\documentclass[fleqn,usenatbib]{mnras}
\usepackage{newtxtext,newtxmath}
\usepackage[T1]{fontenc}
\usepackage{graphicx}	
\usepackage{amsmath}	
\usepackage[dvipsnames]{xcolor}

\defcitealias{Russell2015}{HRR15}

\begin{document}

\title[Feeding and Feedback at the Bondi Radius of M84]{AGN Feeding and Feedback in M84: From Kiloparsec Scales to the \\ Bondi Radius}

\author[C.J. Bambic et al.]  
    {\parbox[]{7.in}{C.~J. Bambic$^{1}$\thanks{E-mail: 
          cbambic@princeton.edu}, H.~R. Russell$^2$, C.~S. Reynolds$^{3,4,5}$,
          A.~C. Fabian$^{3}$, B.~R. McNamara$^{6,7}$ and P.~E.~J. Nulsen$^{8,9}$ \\
    \footnotesize 
    $^1$ Department of Astrophysical Sciences, Peyton Hall, Princeton University, Princeton, NJ 08544, USA \\
    $^2$ School of Physics, Astronomy, University of Nottingham, University Park, Nottingham NG7 2RD, UK \\
    $^3$ Institute of Astronomy, Madingley Road, Cambridge CB3 0HA, UK \\
    $^4$ Department of Astronomy, University of Maryland, College Park, MD 20742-2421, USA \\
    $^5$ Joint Space Science Institute (JSI), College Park, MD 20742-2421, USA \\
    $^6$ Department of Physics and Astronomy, University of Waterloo, 200 University Avenue West, Waterloo, ON N2L 3G1, Canada \\
    $^7$ Waterloo Centre for Astrophysics, University of Waterloo, Waterloo, ON N2L 3G1, Canada \\
    $^8$ Harvard-Smithsonian Center for Astrophysics, 60 Garden Street, Cambridge, MA 02138, USA \\
    $^9$ ICRAR, University of Western Australia, 35 Stirling Hwy, Crawley, WA 6009, Australia
  }
}

\maketitle

\begin{abstract}
We present the deepest \textit{Chandra} observation to date of the galaxy M84 in the Virgo Cluster, with over 840 kiloseconds of~data provided by legacy observations and a recent 730 kilosecond campaign. The increased signal-to-noise allows us to study the origins of the accretion flow feeding the supermassive black hole in the center of M84 from the kiloparsec scales of the X-ray halo to the Bondi radius, $R_{\rm B}$. Temperature, metallicity, and deprojected density profiles are obtained in four sectors about M84's AGN, extending into the Bondi radius. Rather than being dictated by the potential of the black hole, the accretion flow is strongly influenced by the AGN's bipolar radio jets. Along the jet axis, the density profile is consistent with $n_e \propto r^{-1}$; however, the profiles flatten perpendicular to the jet. Radio jets produce a significant asymmetry in the flow, violating a key assumption of Bondi accretion. Temperature in the inner kiloparsec is approximately constant, with only a slight increase from 0.6 to 0.7 keV approaching $R_{\rm B}$, and there is no evidence for a temperature rise imposed by the black hole. The Bondi accretion rate $\dot{M}_{\rm B}$ exceeds the rate inferred from AGN luminosity and jet power by over four orders of magnitude. In sectors perpendicular to the jet, $\dot{M}_{\rm B}$ measurements agree; however, the accretion rate is $> 4 \sigma$ lower in the North sector along the jet, likely due to cavities in the X-ray gas. Our measurements provide unique insight into the fueling of AGN responsible for radio mode feedback in galaxy clusters.
\end{abstract}


\section{Introduction}
\label{sec:intro}

Accretion onto active galactic nuclei (AGN) at the centers of massive elliptical galaxies fuels AGN feedback in clusters of galaxies. The gravitational potential energy released from plasma flowing onto these supermassive black holes (SMBHs) powers jets of relativistic particles which sculpt the surrounding intracluster medium (ICM). In cool-core clusters where the central cooling time of the hot ($10^7-10^8$~K) ICM is $\lesssim$ Gyr, thermalization of the jet kinetic energy provides the heating necessary to balance radiative cooling, maintaining cluster atmospheres in their observed quasi-thermal equilibrium and averting a ``cooling catastrophe'' \citep{Fabian1994,McNamara2007,Fabian2012}.

Deep ($\gtrsim$100 kiloseconds), spatially-resolved \textit{Chandra} X-ray Space Telescope observations of nearby galaxy clusters such as Perseus and Virgo have revealed the signatures of this feedback process: cavities or bubbles carved out of the ICM by jets \citep{McNamara2000,Churazov2001}; weak shocks, ripples, and waves emanating from newly formed bubbles \citep{Sanders2007,Sanders2008,Forman2007}; bright filaments formed by gas cooling around these cavities \citep{Fabian2003}; and turbulent fluctuations stirred by the buoyant rise of bubbles through clusters \citep{Churazov2004,Zhuravleva2014,Hitomi2016,Simionescu2019}. Yet, while the X-ray morphology of clusters has provided insight into how AGN shape their environments on $10\rm{s}-100\rm{s}$ of kiloparsecs (kpc) scales, understanding the connection between AGN and the sub-kiloparsec scale accretion flows which power them remains a critical uncertainty in the paradigm of AGN feedback. 

Even the basic energetics of large-scale black hole ``feeding'' is an open problem (see \citealt{Abramowicz2013} for a review). In the standard paradigm, material within the SMBH's sphere of influence, the Bondi radius ($R_{\rm B} = 2 G M_{\rm BH}/c_{s}^2$ where $M_{\rm BH}$ is the black hole mass and $c_s$ is the speed of sound well beyond $R_B$), is destined to either reach the hole or race away in an outflow. Ionized gas pierces the sphere of influence at a rate $\dot{M}_{\rm B}$ en route to the hole, where this plasma is consumed at a rate $\dot{M}$. A fraction $\eta$ of the rest mass power $\dot{M}c^2$ is released by the accretion flow in the form of radiation and outflows---winds or jets---such that the total power (radiative + outflow) of the AGN is $L = \eta \dot{M} c^2$. 

At the largest scales, a gas inflow with accretion rate $\dot{M}_{\rm B}$ is formed by gas cooling and gravitational infall under the influence of the combined galactic and SMBH potential \citep{Quataert2000}. The accretion rate $\dot{M}$ is then influenced by the structure of this inflow: the angular momentum \citep{Proga2003} and effective turbulent viscosity \citep{Narayan2011} of the gas, and the relative contributions of hot X-ray emitting plasma \citep{DiMatteo2003} vs. cold atomic and molecular gas \citep{Pizzolato2005} which may ``rain down'' through the Bondi radius \citep{Gaspari2012,Yang2016}. Magnetic fields certainly complicate this picture, with the magnetic flux frozen into the flow \citep{Lubow1994} competing with dynamo-generated fields \citep{Brandenburg1995,Brandenburg2005,Blackman2012,Liska2020} to power relativistic jets \citep{Blandford1977,Komissarov2001,Tchekhovskoy2010,Tchekhovskoy2011} and winds \citep{Blandford1982,Proga2000}, and thereby influence the value of $\eta$. 

The complexities of this inflow determine the state of the resulting accretion disk around the black hole and the relative contribution of radiation to the flow's structure. For the jetted systems of interest in cluster AGN feedback, the accretion flow is likely radiatively inefficient, forming a virialized, geometrically-thick advection-dominated accretion flow \citep[ADAF;][]{Ichimaru1977,Rees1982,Narayan1994,Narayan1995,Quataert1999_ADAF}, a convection-dominated accretion flow \citep[CDAF;][]{Quataert2000_CDAF}, or when the net magnetic flux reaching the hole is large, a magnetically arrested disk \citep[MAD;][]{Bisnovatyi1974,Narayan2003,Igumenshchev2008,McKinney2012,Avara2016,Marshall2018,Ripperda2022}.

While $\dot{M}_{\rm B}$ is crucial in determining accretion flow structure, measuring this parameter is a major challenge. Because the true $\dot{M}_{\rm B}$ cannot be measured directly, large-scale black hole feeding is often interpreted through a steady, spherically symmetric model of accretion, the \cite{Bondi1952} solution. Within this framework, $\dot{M}_{\rm B}$ for a given SMBH mass is specified entirely by the gas density and temperature at $R_{\rm B}$, quantities which in principle can be measured with deep X-ray observations. 

This choice is one of convenience---there are no strong theoretical reasons to expect the assumptions of the Bondi solution to hold in real systems. However, some evidence points to the importance of $\dot{M}_{\rm B}$ in setting feedback power. \cite{Allen2006}, using a small sample of X-ray observations of nearby elliptical galaxies, found an apparent correlation between Bondi accretion rate and AGN jet power, as measured from the enthalpy of jet-blown cavities. This method for inferring jet power is subject to significant uncertainties, e.g. projection effects and the assumption of subsonic inflation. Indeed, a follow-up study by \cite{Russell2013} using a larger sample of elliptical galaxies found a less significant correlation. 

A direct correlation between Bondi accretion rate and AGN feedback power has interesting consequences. The correlation may imply a universality in the radiatively inefficient accretion flows (RIAFs) which power AGN in early type galaxies, with $\dot{M}_{\rm B}$ serving as the crucial parameter for regulating power from radiative ($L_{\rm Rad}$) and jet ($L_{\rm Jet}$) feedback on $\sim$Gyr timescales. In addition, the correlation could be leveraged in sub-grid models for galaxy formation, where feedback power from unresolved AGN must be tuned based on resolvable properties, such as $\dot{M}_{\rm B}$ \citep{Pillepich2018}. Establishing this correlation necessitates deep X-ray observations which resolve the density and temperature at $R_{\rm B}$.

In this paper, we harness the deepest X-ray observations to date of the galaxy M84 (NGC 4374) to measure the Bondi accretion rate of hot phase ($\gtrsim$ 0.5 keV) gas onto a jetted AGN in an early type galaxy. These measurements are based on a new \textit{Chandra} campaign which yielded approximately 730 kiloseconds (ks) on M84. Combined with legacy data published in \cite{Finoguenov2001,Finoguenov2002} and \cite{Finoguenov2008}, the observations presented comprise over 840 ks of X-ray data.

M84 is one of only 5 known systems where the Bondi radius can be resolved by \textit{Chandra}, despite the observatory's remarkable sub-arcsecond angular resolution. The other 4 systems are Sgr A$^{*}$ \citep{Baganoff2003}, NGC 3115 \citep{Wong2014}, NGC 1600 \citep{Runge2021}, and M87 \citep[hereafter \citetalias{Russell2015}]{Russell2015}. Even within this small class, M84 stands out. Unlike Sgr~A$^{*}$ and NGC 1600, M84 has an X-ray detected AGN. In contrast to NGC 3115, a \cite{Fanaroff1974} Type I radio jet is clearly observed in M84. However, unlike that in M87 which hosts a notably powerful jet, M84's AGN is not particularly luminous (more than an order of magnitude dimmer than M87's AGN) and our extended campaign caught the SMBH in a relatively quiescent state. Thus, M84 does not require the same sophisticated treatment of pile-up as was performed in M87 \citepalias{Russell2015}. These factors make M84 an especially useful object for exploring the interplay of feeding and feedback in elliptical galaxies. 

This paper is organized as follows. We describe our data analysis in \S\ref{sec:data_analysis} including data reduction, spectral models for the AGN and galactic gas, and simulations of the detector point spread function (PSF) used for forward modelling spectral contamination from the AGN. In \S\ref{sec:results}, we present results: profiles of gas density, temperature, and metallicity approaching and just within the Bondi radius, and the measured Bondi accretion rates $\dot{M}_{\rm B}$ and efficiencies~$\eta$. We discuss the implications of our measurements in \S\ref{sec:discussion}, and conclude in \S\ref{sec:conclusion}. 

\section{\textit{Chandra} Data Analysis} \label{sec:data_analysis}

M84 is a nearby (luminosity distance $D_L=$ 16.83 Mpc; redshift $z=$ 0.00327) giant  elliptical  galaxy  (type  E1) and satellite member of the Virgo Cluster of galaxies. The galaxy has been the subject of three separate \textit{Chandra} ACIS-S campaigns which together yield $\approx$840 ks of data. While earlier works by \cite{Finoguenov2001} and \cite{Finoguenov2008} addressed the detailed structure of M84 and how the X-ray halo is shaped by feedback, our ultra-deep campaign is concerned primarily with black hole feeding and gas structure approaching and just within the Bondi radius of the SMBH.

\subsection{Data Reduction} \label{sec:data_reduction}

This work is a follow-up to a similar analysis of M87 by \citetalias{Russell2015}. Thus, we follow the same data reduction procedure. 

\begin{figure*}
\hbox{
\includegraphics[width=0.5\textwidth]{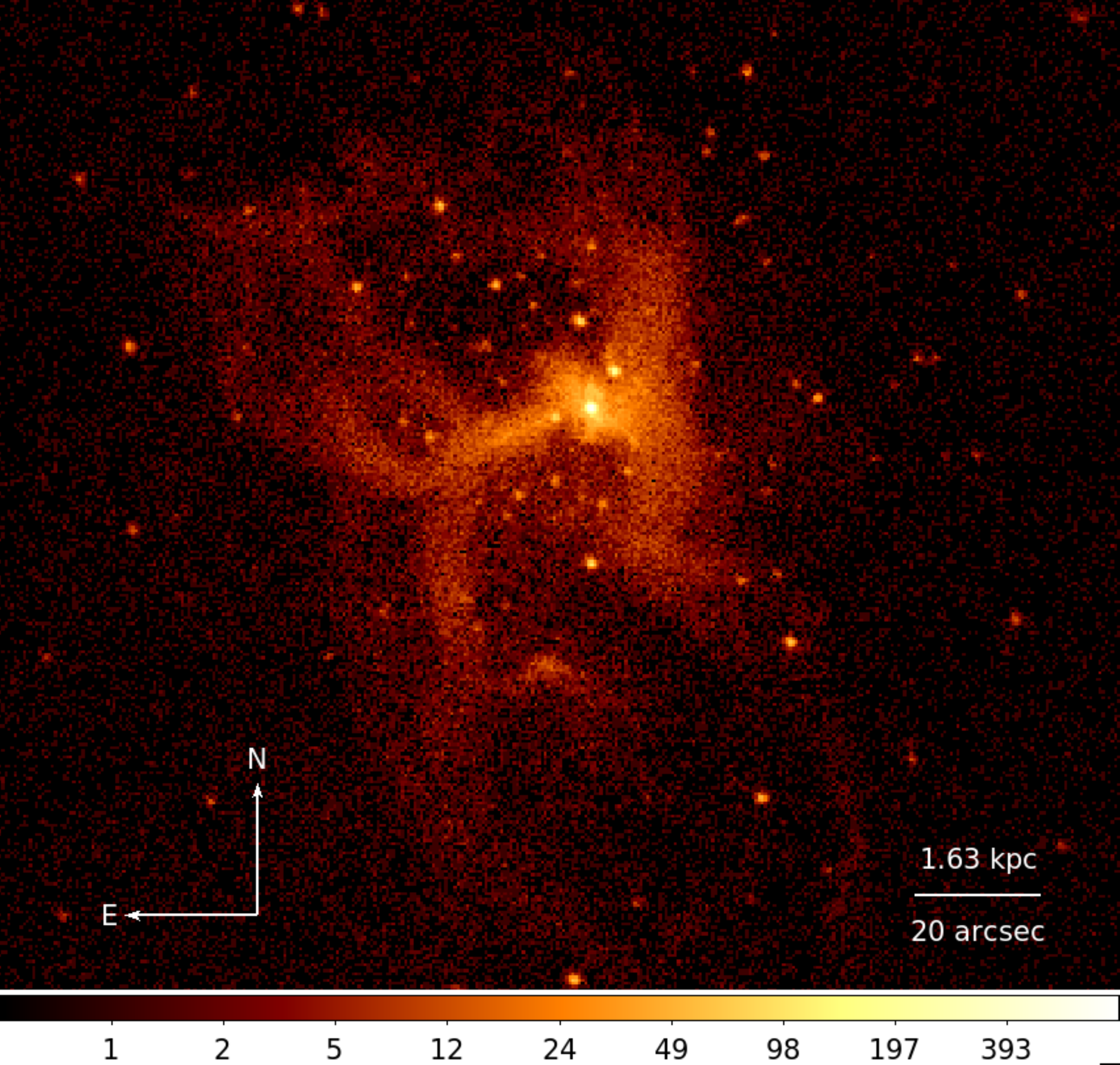}
\includegraphics[width=0.5\textwidth]{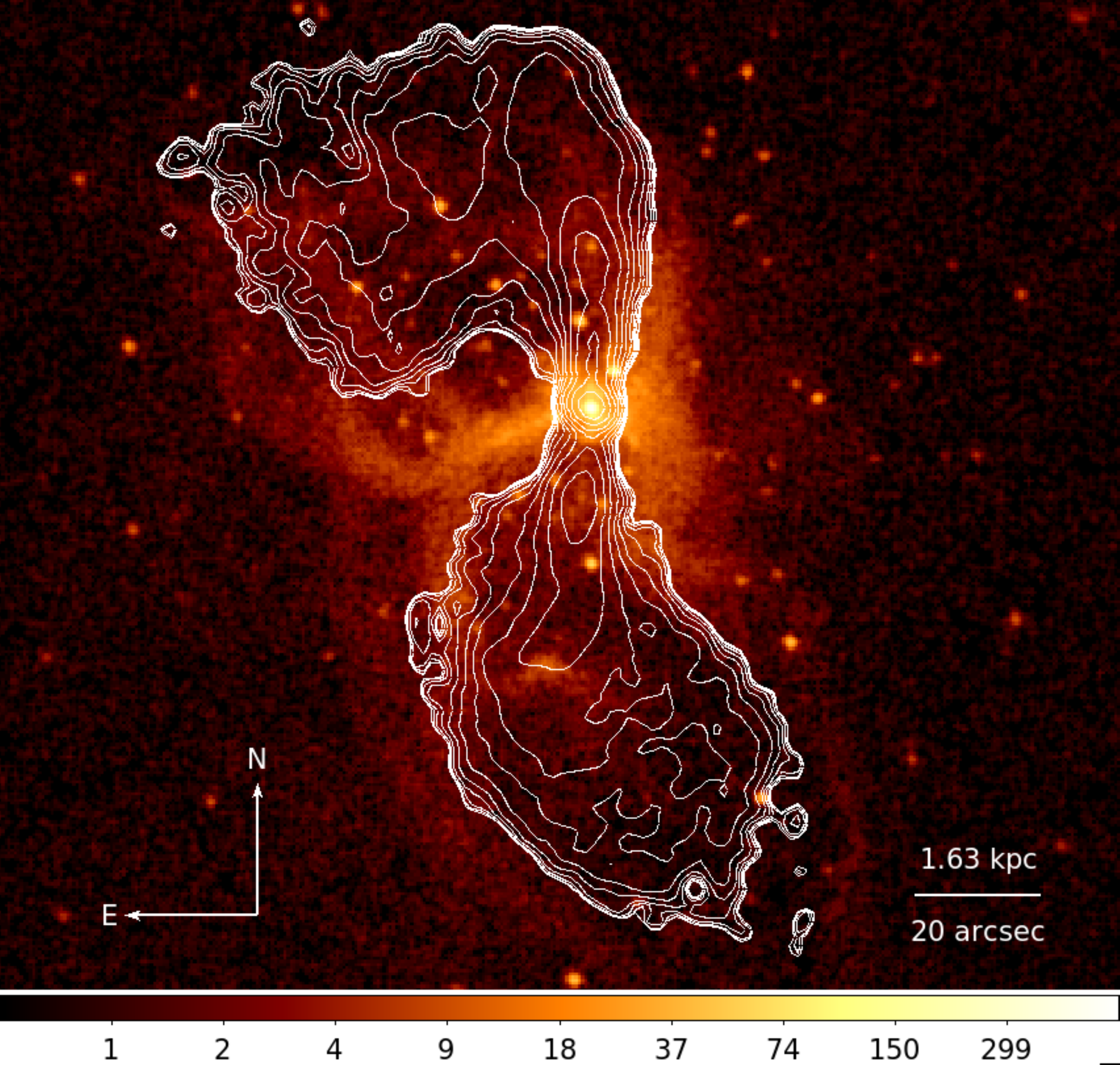}
}
\caption{Left: Merged $0.5-2$ keV image, totaling 798.66 ks of cleaned exposure time with Obs. ID 5908 excluded (see \S\ref{sec:variability}). Right: 5 GHz radio contours (white) as measured by the Very Large Array (VLA) overlaid on X-ray image. Contours correspond to 12 logarithmically-spaced levels in flux from $2 \times 10^{-3} - 0.1$~mJy, and colors denote X-ray counts. The $\mathcal{H}$-shaped morphology is carved out of the halo by radio jets, forming bright rims about the radio lobes.
}
\label{fig:beautiful_image}
\end{figure*}

Our data reduction was performed using \texttt{CIAO} version 4.11 and the Calibration Database (CalDB) 4.8.5, updated November 7, 2019 \citep{Fruscione2006}. This update followed a major revision to the soft energy response brought about by contaminant build-up over \textit{Chandra}'s prolific 23 year lifetime (thus far). Our long campaign was affected by this contamination, and as we show, the majority of M84's galactic gas, especially that approaching the Bondi radius, is cooler than 1 keV and emitting X-rays within the range of degraded performance. Given the low temperature of the extended emission in M84, the calibration of the contaminant build up on Chandra's optical block filters is particularly important.  We therefore verified that temperature, metallicity and normalization values measured with the new observations are consistent with the archival observations, which were taken only a few years after \textit{Chandra}'s launch and less affected. Using the \texttt{chandra\_repro} routine, we reprocess our data to produce second-level event files, removing bad pixels based on the analysis reference data library (\texttt{ardlib}), detecting point sources using \texttt{wavdetect}, and creating light curves to filter out bad time intervals. To produce merged images, we assume an exposure correction for each Obs. ID's exposure map.

\subsection{X-ray Morphology} \label{sec:morphology}

Figure~\ref{fig:beautiful_image} displays a merged $0.5-2$ keV image based on all three campaigns. Similar images can be found in \cite{Finoguenov2001} and \cite{Finoguenov2008} from the first two sets of observations.

Using only limited \textit{Chandra} data, \cite{Finoguenov2001} were able to identify the salient features of the galaxy's X-ray emission. Instead of a featureless X-ray halo, M84 hosts depressions in emissivity North and South of the central AGN, coincident with radio lobes produced by \cite{Fanaroff1974} Type I jet activity \citep{Laing1987}. These cavities create an $\mathcal{H}$-shaped structure in the halo gas, which extends $\approx 150''$ (12.2 kpc) from the Northernmost edge of the emission to the faint rim in the Southwest of the image. The crossbar of the $\mathcal{H}$ spans $\approx 46''$ (3.7 kpc) and is approximately aligned with optical dust lanes \citep{Hansen1985}, although the dust lanes are on a much larger scale, cutting across the X-ray image. 

As argued by \cite{Finoguenov2008}, these cavities may actually be comprised of at least two bubbles each, with bright rims (viewed in projection) demarcating the bubble boundaries. Indeed, our deep observation is able to clearly detect a tenuous bubble rim extending toward the Southwest in the image. The Northern bubble is compressed, likely by the ram pressure of ICM gas as the galaxy moves through the cluster. 

While the bubbles are located just to the North and South of the crossbar, the jet is aligned with the West filament; the galaxy has drifted over time. Subsequently, ram pressure has swept the Northern bubble back and ``bent'' the radio jet---a signature of radio galaxies moving through clusters \citep{Miley1972, Owen1976, Begelman1979, Morsony2013, McBride2014}. Intriguingly, the Southern bubble has not been swept in the same direction. There may be a large-scale shear flow across M84, or the jet may have reoriented itself over the course of the episodes recorded in the radio lobes, possibly through precession. 

\begin{table*} 
\renewcommand{\arraystretch}{1.1}
\small\addtolength{\tabcolsep}{-2pt}
\scalebox{1}{%
\begin{tabular}{c c c c c c c}     
\hline  
Obs. ID & Date & Exposure  & $N_{\mathrm{H}}$ & $\Gamma$ & Flux (2-10 keV) & C-Stat/DOF \\ 
        &      &  (ks)     &(10$^{22}$ cm$^{-2}$)&  & (10$^{-13}$ erg cm$^{-2}$ s$^{-1}$) & \\  
\hline
803   & 19/05/2000 & 28.47 & 0.23$_{-0.07}^{+0.09}$ & 1.79$_{-0.19}^{+0.20}$ & 0.99$_{-0.12}^{+0.14}$ & 142.6/ 184\\
5908   & 01/05/2005 & 46.08 & 0.16$_{-0.03}^{+0.03}$ & 2.03$_{-0.10}^{+0.10}$ & 1.62$_{-0.11}^{+0.12}$ & 233.7/ 277\\
6131   & 07/11/2005 & 40.93 & 0.81$_{-0.31}^{+0.39}$ & 1.68$_{-0.34}^{+0.36}$ & 0.63$_{-0.08}^{+0.10}$ & 146.3/ 165\\
20539   & 05/04/2019 & 39.54 & 0.16$_{-0.16}^{+0.29}$ & 1.66$_{-0.31}^{+0.36}$ & 0.50$_{-0.07}^{+0.09}$ & 121.8/ 141\\
20540   & 26/02/2019 & 30.17 & 0.08$_{-0.08}^{+0.28}$ & 1.78$_{-0.25}^{+0.38}$ & 0.49$_{-0.08}^{+0.09}$ & 109.5/ 127\\
20541   & 10/04/2019 & 11.29 & 0.37$_{-0.37}^{+0.66}$ & 2.11$_{-0.80}^{+0.97}$ & 0.46$_{-0.15}^{+0.26}$ & 42.1/ 55\\
20542   & 18/03/2019 & 34.61 & 0.005$_{-0.005}^{+0.28}$ & 1.46$_{-0.18}^{+0.36}$ & 0.48$_{-0.08}^{+0.08}$ & 120.9/ 122\\
20543   & 27/04/2019 & 54.34 & 1.57$_{-0.52}^{+0.63}$ & 2.95$_{-0.48}^{+0.53}$ & 0.32$_{-0.04}^{+0.05}$ & 113.5/ 135\\
21845   & 28/03/2019 & 27.70 & 0.50$_{-0.32}^{+0.25}$ & 2.05$_{-0.42}^{+0.34}$ & 0.47$_{-0.08}^{+0.09}$ & 118.8/ 113\\
21867   & 13/03/2019 & 23.63 & 0.36$_{-0.35}^{+0.53}$ & 2.31$_{-0.52}^{+0.62}$ & 0.34$_{-0.07}^{+0.08}$ & 80.0/ 104\\
22126   & 28/02/2019 & 35.10 & 0.22$_{-0.16}^{+0.20}$ & 1.80$_{-0.27}^{+0.29}$ & 0.67$_{-0.09}^{+0.11}$ & 132.3/ 153\\
22127   & 02/03/2019 & 22.77 & 0.27$_{-0.21}^{+0.26}$ & 1.75$_{-0.31}^{+0.32}$ & 0.85$_{-0.12}^{+0.14}$ & 93.2/ 136\\
22128   & 03/03/2019 & 23.75 & 0.37$_{-0.25}^{+0.32}$ & 1.78$_{-0.36}^{+0.40}$ & 0.70$_{-0.11}^{+0.14}$ & 94.6/ 124\\
22142   & 14/03/2019 & 20.77 & 0.74$_{-0.47}^{+0.64}$ & 2.53$_{-0.66}^{+0.74}$ & 0.35$_{-0.08}^{+0.10}$ & 58.5/ 86\\
22143   & 16/03/2019 & 22.75 & 0.59$_{-0.34}^{+0.42}$ & 2.05$_{-0.42}^{+0.45}$ & 0.70$_{-0.11}^{+0.13}$ & 111.0/ 123\\
22144   & 15/03/2019 & 31.75 & 0.09$_{-0.09}^{+0.19}$ & 2.03$_{-0.26}^{+0.31}$ & 0.46$_{-0.07}^{+0.08}$ & 128.0/ 142\\
22153   & 23/03/2019 & 21.08 & 0.89$_{-0.41}^{+0.51}$ & 2.58$_{-0.51}^{+0.56}$ & 0.45$_{-0.08}^{+0.10}$ & 74.2/ 93\\
22163   & 29/03/2019 & 35.59 & 0.54$_{-0.27}^{+0.33}$ & 1.86$_{-0.32}^{+0.34}$ & 0.65$_{-0.08}^{+0.10}$ & 125.3/ 153\\
22164   & 31/03/2019 & 32.63 & 0.59$_{-0.39}^{+0.42}$ & 1.85$_{-0.41}^{+0.40}$ & 0.60$_{-0.08}^{+0.10}$ & 134.7/ 142\\
22166   & 06/04/2019 & 38.56 & 0.41$_{-0.24}^{+0.31}$ & 2.31$_{-0.35}^{+0.39}$ & 0.34$_{-0.05}^{+0.06}$ & 117.2/ 132\\
22174   & 11/04/2019 & 49.41 & 0.84$_{-0.30}^{+0.36}$ & 2.20$_{-0.35}^{+0.38}$ & 0.52$_{-0.07}^{+0.08}$ & 129.4/ 160\\
22175   & 12/04/2019 & 27.20 & 0.49$_{-0.34}^{+0.41}$ & 1.83$_{-0.40}^{+0.43}$ & 0.54$_{-0.09}^{+0.11}$ & 93.0/ 114\\
22176   & 13/04/2019 & 51.39 & 0.57$_{-0.20}^{+0.23}$ & 2.21$_{-0.25}^{+0.26}$ & 0.64$_{-0.06}^{+0.07}$ & 165.3/ 195\\
22177   & 14/04/2019 & 36.58 & 0.86$_{-0.32}^{+0.37}$ & 2.59$_{-0.38}^{+0.40}$ & 0.49$_{-0.06}^{+0.07}$ & 100.9/ 147\\
22195   & 28/04/2019 & 38.07 & 0.66$_{-0.49}^{+0.61}$ & 1.98$_{-0.50}^{+0.54}$ & 0.41$_{-0.06}^{+0.08}$ & 133.0/ 122\\
22196   & 07/05/2019 & 20.58 & 0.97$_{-0.44}^{+0.56}$ & 2.72$_{-0.57}^{+0.63}$ & 0.34$_{-0.07}^{+0.08}$ & 67.1/ 88\\
New Campaign & 02/2019-05/2019 & 729.26 & 0.49$_{-0.07}^{+0.07}$ & 2.05$_{-0.08}^{+0.08}$ & 0.50$_{-0.02}^{+0.02}$ & 2645.8/ 2973\\
All Data & 05/2000-05/2019 & 844.74 & 0.44$_{-0.06}^{+0.06}$ & 2.00$_{-0.08}^{+0.08}$ & 0.52$_{-0.02}^{+0.02}$ & 3055.9/ 3328\\
\hline
\end{tabular}}
\caption{Summary of observations. Fits to AGN are obtained using the ``M84 Model'' as presented in the text, with a \texttt{VAPEC} model for the galactic emission and an \texttt{APEC} model for the Virgo Cluster ``screen.'' The photon index $\Gamma$ remains close to 2 as expected for Comptonized emission from an ADAF. We include local absorption with column density $N_{\rm H}$ to account for intervening dust lanes or a dusty torus around the AGN.} 
\label{table:observations} 
\end{table*}

\subsection{Spectral Fitting} \label{sec:spectral_fitting}

Unless stated otherwise, we fit spectra simultaneously with the \texttt{XSPEC} spectral fitting package \citep{Arnaud1996} using all Obs. IDs listed in Table~\ref{table:observations}. Spectra are extracted using \texttt{CIAO}'s \texttt{specextract} function and grouped such that at least one count is present in all energy bins over the range of $0.5-7$ keV. Fits are performed using the modified C-statistic \citep{Cash1979} with elemental abundances taken from \cite{Anders1989} for comparison with past results. All spectral models are fit with galactic absorption included via a photoelectric absorption (\texttt{phabs}) model, with a constant galactic column density of $N_{\rm H, gal}~=~2.9~\times~10^{20}$~cm$^{2}$ as measured by the HI4Pi Survey \citep{HI4PI2016}. 

\subsection{Virgo Cluster Spectral Model} \label{sec:virgo_model}

M84 is embedded in the Virgo Cluster, so we must peer through a ``screen'' of hard X-ray emission $\gtrsim$1~keV. Because the Virgo Cluster occupies the entire field of view, we follow \citetalias{Russell2015} and choose to use blank sky backgrounds for all spectral fits. The appropriate blank-sky background dataset was processed identically to the event file, reprojected to the same sky position, and normalized so that the count rate matched that of the event file for the $9.5-12$ keV energy band.

We model the bremsstrahlung emission from the Virgo ICM with a single temperature \texttt{APEC} plasma emission model. The parameters for this model are determined by fitting a spectrum extracted from a large region, a $4.6' \times 4.9'$ box around M84, excluding point sources (as detected by \texttt{wavdetect}) and the galaxy, whose X-ray emission is confined within a $1.9' \times 2.6'$ box. 

This Virgo spectrum is fit with temperature ($T$), metallicity ($Z$), and normalization free. The fit yields reasonable values, $T$~=~2.32~$\pm$~0.06~keV and $Z$~=~0.429~$\pm$~0.04~$Z_{\odot}$,  consistent with $T\approx$ 2.3 keV obtained by \cite{Urban2011} and \cite{Ehlert2013} using the much higher spectral resolution of XMM-\textit{Newton}.

\begin{figure*}
\hbox{
\includegraphics[width=1.0\textwidth]{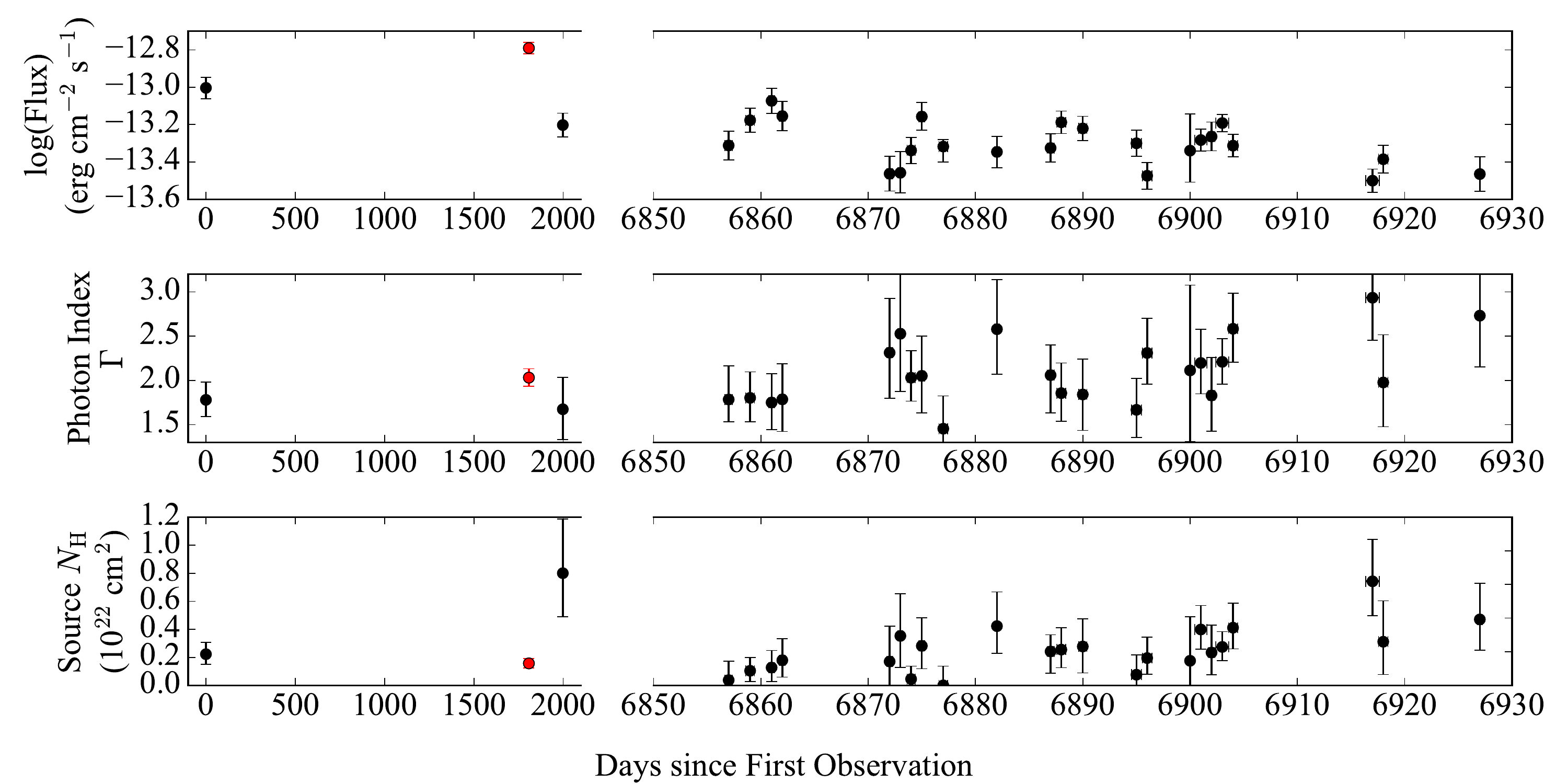}
}
\caption{Time variability of AGN over all three observational campaigns used in this work. The observation corresponding to Obs. ID 5908 (red) has a much higher flux than the others, violating an assumption that the AGN has a constant luminosity. We remove this observation from our analysis of Bondi radius scales.
}
\label{fig:AGN_Variability}
\end{figure*}

\subsection{M84 Galactic Gas Spectral Model} \label{sec:galactic_model}

The earliest \textit{Chandra} measurements of M84 by \cite{Finoguenov2001} showed an overabundance of metals relative to solar. This overabundance could be contributed both by iron-peak elements (Fe~and~Ni) originating from Type Ia supernovae, or $\alpha$ elements (C, N, O, Al, Si, etc.) produced by Type II supernovae. Since XMM-\textit{Newton} lacks the spatial resolution of \textit{Chandra}, abundance measurements performed by XMM probe larger length scales than we are studying; we must constrain $\alpha$ element metallicities ourselves. 

We fit the spectrum of the full $4.6' \times 4.9'$ box with the galaxy included using a \texttt{VAPEC} model for M84's galactic gas emission, an \texttt{APEC} component for the Virgo ICM, and an extra power law component for unresolved point sources (see \S\ref{sec:unresolved_points}). Since the helium (He) abundance of the \texttt{VAPEC} component cannot be constrained in the X-ray band, we set the He abundance to solar. \texttt{VAPEC} iron-peak element abundances $Z_{\rm Fe}$ are tethered together in the fits, as are all remaining $\alpha$ element metallicities $Z_{\alpha}$. The \texttt{APEC} temperature and metallicity are fixed based on \S\ref{sec:virgo_model}, but the normalization is left free.

\subsection{Unresolved Point Sources} \label{sec:unresolved_points}

M84 is known to host a substantial number of X-ray binary (XRB) point sources \citep{Finoguenov2002}. While many of these XRBs can be masked out, unresolved XRBs and AB/CV stars represent a source of hard emission which can affect temperature and abundance measurements. We follow the common practice of modeling these unknown populations using a simple power law model with fixed photon index $\Gamma_{\rm XRB} = 1.6$ \citep{Goulding2016}. The normalization of this power law is left free in the fits to the \texttt{VAPEC}+\texttt{APEC} model, which are designed to provide adequate statistics for constraining the $\alpha$ element metallicity, $Z_{\alpha}$.

Unfortunately, the contaminant build-up which has degraded \textit{Chandra}'s soft energy response prevents us from constraining the $\alpha$ element metallicity from the new extended campaign, even with ample source counts available. Thus, the only Obs. IDs used for determining $Z_{\alpha}$ come from legacy observations: Obs. IDs 803, 5908, and 6131. We find a reasonable constraint on $Z_{\alpha}$, approximately 0.45 times solar abundance. This value changes only slightly to 0.51 solar when the power law component is neglected. For all remaining spectral fits in this paper, we fix $Z_{\alpha}$ to 0.45 and fit only the temperature, normalization, and the iron-peak metallicities in the \texttt{VAPEC} model.

\subsection{Accounting for AGN Contamination} \label{sec:AGN_contamination}

While the AGN is a point source, \textit{Chandra}'s point spread function (PSF) distributes AGN photons across a number of pixels, including those containing photons produced by gas at Bondi radius scales. Therefore, we must account for AGN contamination when fitting spectra extracted from these sub-kpc scales. 

The AGN and galactic gas are spectrally \textit{distinct}; however, they are not spectrally separable with \textit{Chandra}'s spectral resolution. We find that when fitting the AGN and galactic gas together, we are unable to adequately constrain the parameters for both models. In response to this limitation, we follow the example of \citetalias{Russell2015} and forward-model the AGN contamination by simulating \textit{Chandra}'s energy-dependent PSF's effect on the measured AGN spectrum. We first fit the spectrum of the AGN based on fixing the parameters of the galactic gas spectral model (\S\ref{sec:galactic_model}). Using the parameters obtained from this fit, we produce an AGN spectrum free of contributions from the galactic gas or the Virgo screen. This spectrum is fed into the \texttt{Cha}ndra \texttt{R}ay \texttt{T}racer tool \citep[\texttt{ChaRT};][]{Carter2003_ChaRT} which simulates the detector response. The \texttt{MARX} software package \citep{Davis2012_MARX} is used to produce a second-level event file of the simulated AGN.

\begin{figure*}
\hbox{
\includegraphics[width=0.45\textwidth]{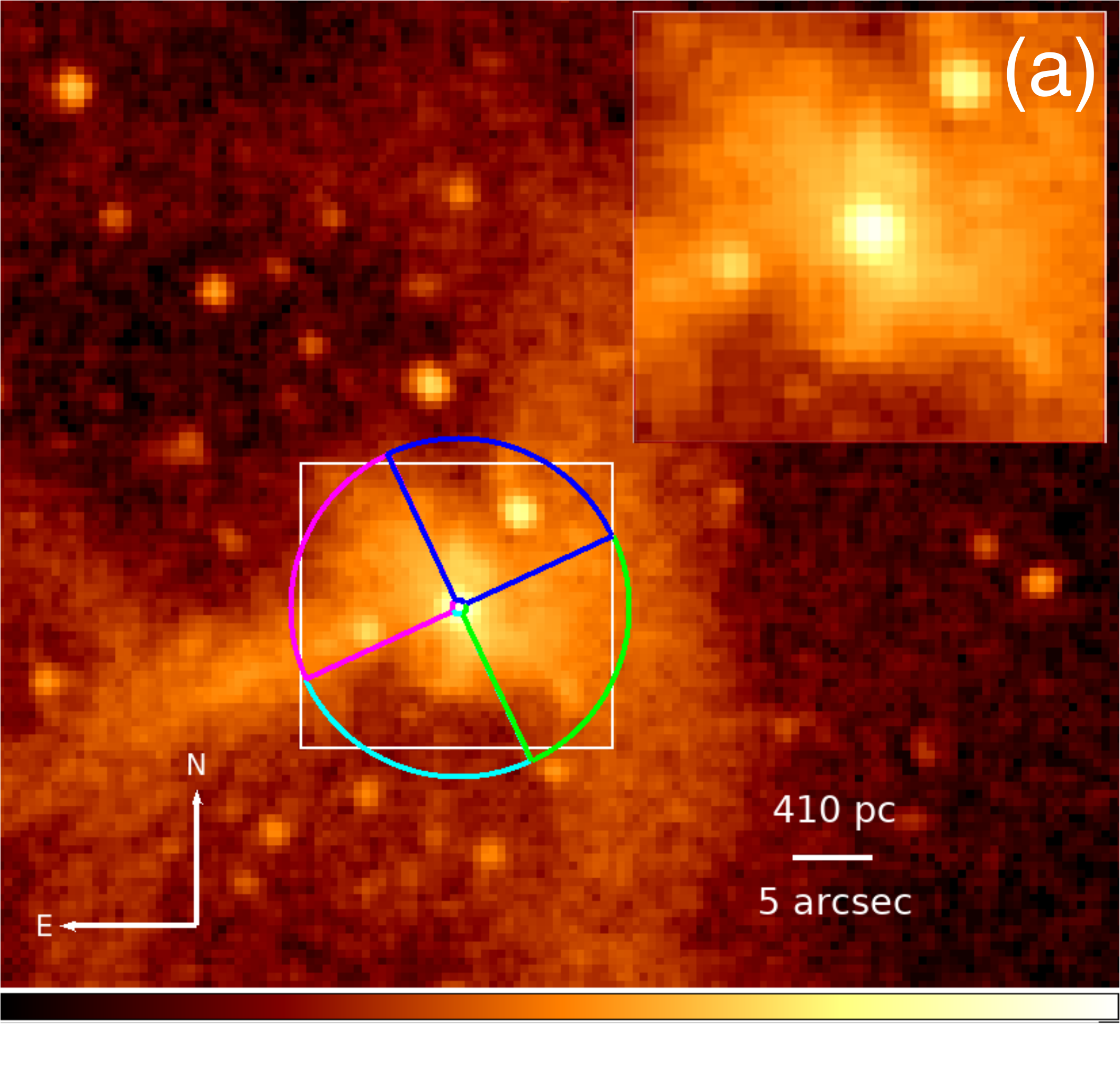}
\includegraphics[width=0.56\textwidth]{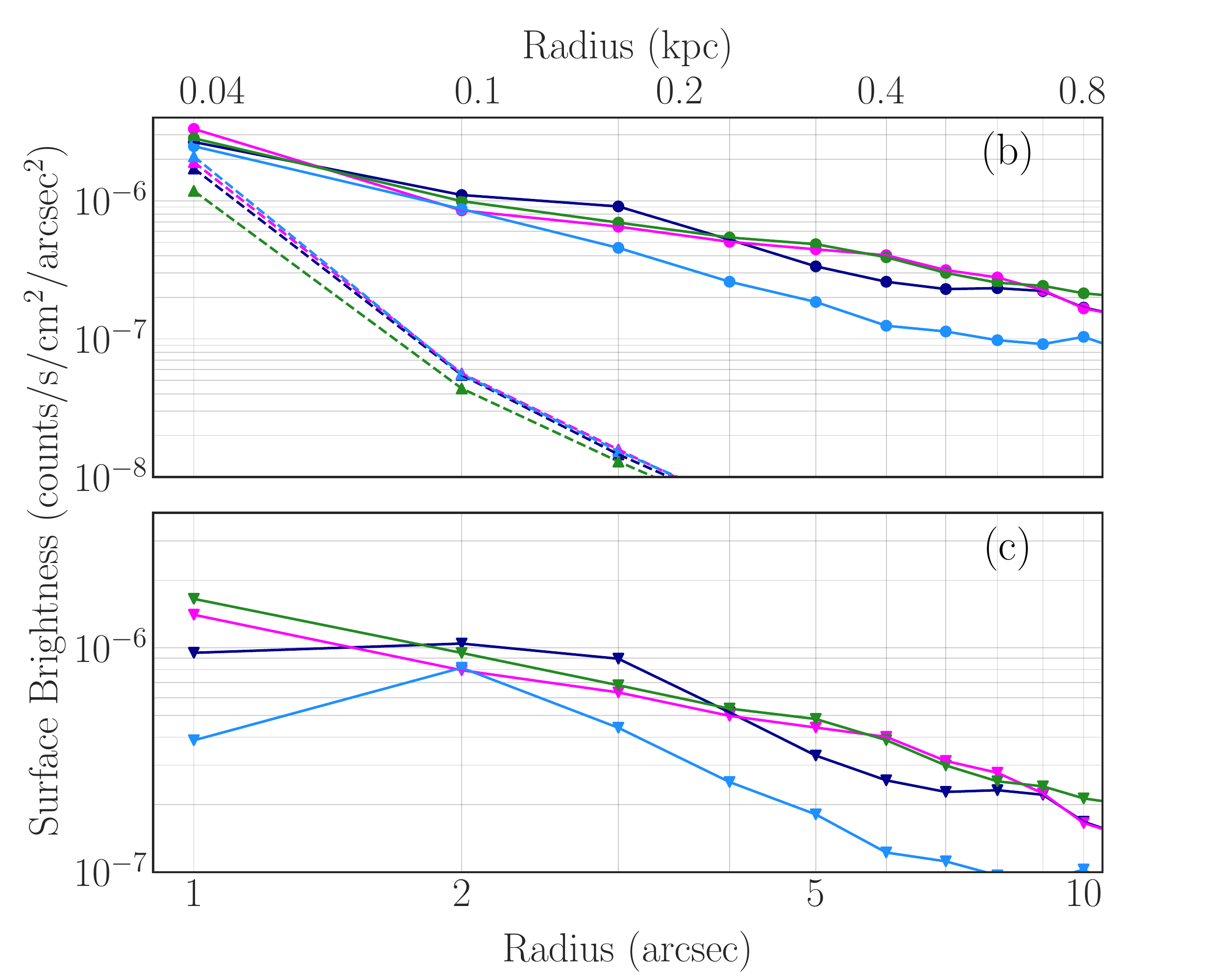}
}
\caption{Left: Sectors ($10''$ in radius) overlaid on merged $0.5-2$ keV image of M84 central region. The East (magenta) and West (green) regions are aligned perpedicular with the AGN jet which is approximately aligned with the Western filament (Figure~\ref{fig:beautiful_image}). Right: Background-subtracted $\mathbf{0.5-7}$ keV surface brightness profiles in sectors. The upper panel shows profiles of the AGN (triangles) as well as the broadband data, while the lower panel shows the AGN-subtracted surface brightness profile, used to compute the deprojected density profiles.}
\label{fig:SB_profile}
\end{figure*}

\subsection{AGN Spectral Model} \label{sec:AGN_model}

Previous estimates based on the most recent SMBH mass measurements by \cite{Walsh2010} place the angular scale of the SMBH Bondi radius at approximately $1''$ \citep{Russell2013}; however, measurements in the literature differ by as much as a factor of $\approx$4 \citep{Bower1998,Maciejewski2001}. Thus, we choose to define the region of AGN contamination as a circle with a radius of $1''$ centered on the peak of the AGN surface brightness distribution. When we refer to the ``AGN spectrum'' in this paper, we are referring to the spectrum extracted from this region which includes the AGN, galactic gas emission, Virgo Cluster emission, and unresolved XRBs and AB/CV stars. 

The Comptonized emission from the AGN is modeled as a redshifted power law \citep{Russell2010,Siemiginowska2010}. Due to the presence of the intervening dust lanes as well as a possible ``dusty torus'' surrounding the AGN, we allow for local photoelectric absorption through a \texttt{zphabs} model. Thus, the spectrum of the $1''$ AGN region is fit using the ``M84 model,'' \texttt{phabs(zphabs(zpowerlw)+VAPEC+APEC+powerlaw)}. While this model seems complicated, the only free parameters are the local column density and the photon index $\Gamma$ and normalization of the \texttt{zpowerlw} component. 

Appendix A2 of \citetalias{Russell2015} presents the method for obtaining the parameters for the remaining components. Spectra are extracted from a $2''-4''$ annulus circumscribing the AGN and fit with a \texttt{phabs(VAPEC+APEC+powerlaw)} model for the galactic gas, Virgo ICM, and unresolved point source emission. The \texttt{APEC} component is fixed based on fits in \S\ref{sec:galactic_model} and the $T$ and $Z$ from \S\ref{sec:virgo_model} (note that all normalizations are scaled to the appropriate region areas). We then fit the annulus spectrum with a free \texttt{VAPEC} temperature, iron-peak metallicity, and normalization as well as a free \texttt{powerlaw} normalization with fixed photon index $\Gamma_{\rm XRB} = 1.6$ (see \S\ref{sec:galactic_model} and \S\ref{sec:unresolved_points}). 

The \texttt{VAPEC} normalization used in the AGN fits is determined by boosting the normalization from the annulus fit centered at $3''$ to the $1''$ AGN circle based on a power law extrapolation of the $1''-40''$ surface brightness profile. For the \texttt{powerlaw} component, we assume the normalization is constant from $1''-3''$.

\subsection{AGN Variability} \label{sec:variability}

We perform the fit described in \S\ref{sec:AGN_model} for all Obs. IDs individually in addition to a simultaneous fit. The measured parameters for the AGN model are shown in Table~\ref{table:observations} and plotted in Figure~\ref{fig:AGN_Variability}. 

Note that the AGN, unlike the galactic gas emission, is highly variable, with nearly an order of magnitude variation in flux over the three campaigns. Serendipitously, the AGN was relatively quiescent during 2019 (observations following the break in Figure~\ref{fig:AGN_Variability}), implying that the vast majority of data is subject to minimal AGN contamination. Unfortunately, one observation, Obs. ID 5908, caught a state of outburst. Because we are modeling the AGN based on the statistically powerful simultaneous fit to \textit{all} usable observations, we necessarily make the assumption that the AGN flux is constant with time. Obs. ID 5908 (shown in red in Figure~\ref{fig:AGN_Variability}) breaks this assumption and is thus omitted from the remaining analysis of Bondi radius scales. 

\subsection{Simulating the AGN Spectrum} \label{sec:chart_marx}

The fit to the AGN spectrum is consistent with a photon index of $\Gamma \approx 2$, in accord with other similar ADAF spectra \citep{Gu2009,Younes2011,Yang2015,Younes2019}. With this result, we simulate how the AGN spectrum free of contributions from the galactic gas, Virgo, and XRB/AB/CVs would appear to \textit{Chandra}'s ACIS-S detector. 

Appendix A3 of \citetalias{Russell2015} describes this process in detail. We boost the normalization of the spectrum input to \texttt{ChaRT} and use the aspect solution file for Obs. ID 20543 (the longest exposure observation). Pile-up is negligible in M84 and not included in our modeling. 

The \texttt{MARX} tool is used to produce a second level event file and exposure map for the simulation, and we reproject the simulation to the coordinate system of the observation. We have accounted for galactic emission in the AGN spectrum by measuring the galactic background from $2''-4''$ and modeling the surface brightness (SB) profile to account for the increase in background into $1''$. However, our choice of model introduces a systematic error which may cause us to inadvertently over or under-subtract the AGN by a small amount---enough to impact the measured temperature. 

Subtraction of the AGN can be tested by comparing the hard band ($4-7$ keV) SB profile of the simulation with that of the data (see Appendix A). Bright, lumpy, soft emission from galactic gas, which may vary on scales of the PSF, is entirely subdominant in the $4-7$ keV band. Instead, AGN, Virgo ICM, and unresolved point source emission dominates. If the AGN is under-subtracted, a hard band excess will manifest itself at the scales of the PSF. However, if the AGN is over-subtracted, there should be a drop in hard emission at small scales. Because the Virgo screen is uniform over these small scales, any discontinuities in the AGN-subtracted SB profile is evidence of spatial variation in the unresolved point source flux. 

Our AGN simulation leaves a modest hard band excess of $7\%$ at PSF scales. We compensate for this excess by boosting the AGN simulation 7\% such that the AGN-subtracted $4-7$ keV band SB profile flattens from $1''-2''$ (see Figure~\ref{fig:hard_band}). Continuity of the AGN-subtracted SB at these small scales indicates that the unresolved point source emission is relatively constant with radius, and our assumption that the XRB/AB/CV normalization is constant with radius obtains.

We extract spectra for both the observations and AGN simulation in $1''$ radial bins extending from $1''-10''$. Simulated spectra are fit using an absorbed, redshifted power law model, and the fit parameters are fixed for the AGN components in the combined ``M84 model'' used in \S\ref{sec:AGN_model}. We compute the error bars on temperature and metallicity for the two points in the innermost $2''$ of each sector by boosting or diminishing the AGN normalization by 5\%, marginalizing over uncertainties in the AGN flux. The AGN normalization is set to 0 beyond the innermost three annuli in each sector since the AGN's PSF is insignificant beyond $3''$.

\begin{table*}
\renewcommand{\arraystretch}{1.1}
\small\addtolength{\tabcolsep}{-2pt}
\scalebox{1}{%
\begin{tabular}{c c c c c c c c c}     
\hline  
Sector & $R_{\rm B}$     & $T (R_{\rm B})$ 
       & $Z (R_{\rm B})$ & $n_e (R_{\rm B})$   
       &  Index          & $\dot{M}_{\rm B}$                 
       & $\eta$          & $L_{\rm B}/ L_{\rm Edd}$ \\ 
       
       & pc              &  keV  
       & Z$_{\odot}$     & cm$^{-3}$  
       & $\alpha$        & $10^{-3}\: \mathrm{M}_{\odot} \mathrm{yr}^{-1}$ & $\times 10^{-6}$& $\times 10^{-4}$ \\
\hline

North & $49.0_{-5.3}^{+6.6}$ & $0.71_{-0.05}^{+0.04}$ & $0.14_{-0.03}^{+0.04}$ & $0.11 \pm 0.09$ & $-1.15 \pm 0.19$ & $1.57_{-1.01}^{+1.03}$ & $17.62_{-7.03}^{+31.64}$ & $0.84_{-0.54}^{+0.56}$ \\
East & $48.4_{-5.4}^{+5.7}$ & $0.72_{-0.04}^{+0.05}$ & $0.09_{-0.02}^{+0.03}$ & $0.45 \pm 0.04$ & $-0.80 \pm 0.12$ & $6.35_{-1.35}^{+1.58}$ & $4.37_{-0.93}^{+1.28}$ & $3.39_{-0.78}^{+0.93}$ \\
West & $42.8_{-5.9}^{+4.6}$ & $0.82_{-0.04}^{+0.09}$ & $0.12_{-0.02}^{+0.02}$ & $0.47 \pm 0.03$ & $-0.93 \pm 0.09$ & $5.52_{-1.31}^{+1.21}$ & $5.03_{-0.98}^{+1.68}$ & $2.94_{-0.75}^{+0.72}$ \\
South & $58.7_{-6.9}^{+8.2}$ & $0.60_{-0.05}^{+0.05}$ & $0.29_{-0.07}^{+0.09}$ & $0.26 \pm 0.02$ & $-1.21 \pm 0.33$ & $\leq 4.88$ & $\geq 5.68$ & $\leq 2.60$ \\
All & $48.1_{-4.7}^{+5.3}$ & $0.73_{-0.02}^{+0.02}$ & $0.15_{-0.02}^{+0.02}$ & $0.27 \pm 0.04$ & $-0.87 \pm 0.09$ & $3.74_{-0.89}^{+1.05}$ & $7.42_{-1.72}^{+2.46}$ & $1.99_{-0.50}^{+0.61}$ \\
\hline
\end{tabular}}
\\
\caption{Summary of measurements at $1''$, density profile index $\alpha$, accretion rates $\dot{M}_{\rm B}$, efficiencies $\eta$, and Eddington ratios $L_{\rm B}/L_{\rm Edd}$ for Bondi accretion in each sector. Because of small-scale cavities evident in the AGN-subtracted SB profiles in the South sector (Figure~\ref{fig:SB_profile}), we are only able to obtain limits on the density $n_e$ and quantities derived from density in this region.}  
\label{table:results} 
\end{table*}

\subsection{Profile Deprojection} \label{sec:deprojection}

To obtain density profiles, we follow \citetalias{Russell2015} and first compute background-subtracted surface brightness (SB) profiles of the inner $1''-10''$ ($\sim0.8$ kpc) in sectors, referred to as North, East, West, and South respectively. The North and South sectors are aligned with the radio jet, while the East and West sectors are anti-aligned. We subtract off the Virgo and X-ray background from the SB profiles based on a measurement of SB taken far away from the galaxy and free of point sources. The sectors and SB profiles are shown in Figure~\ref{fig:SB_profile}.
\newpage
SB is a projection of a 3D distribution of X-ray emission onto a 2D plane. By assuming spherical symmetry, we can deproject each sector's SB profile, ``peeling back'' shells of X-ray emission to determine a volumetric emissivity at each profile radius. Spherical symmetry is a poor assumption for M84's highly-structured $\mathcal{H}$-shape; however, the assumption may obtain more readily around the quasi-spherical halo in the inner $10''$.

We apply the deprojection method of \cite{Kriss1983} to the AGN-subtracted SB profiles, panel (c) in Figure~\ref{fig:SB_profile}. This method only strictly applies when the SB is monotonically increasing inward. Cavities, evident in the significant drop in the AGN-subtracted surface brightness profiles for the North and South sectors at a radius of $1''$, violate this assumption and prevent us from obtaining anything more than an upper limit on $n_e$ in the South sector. 

Even though the AGN simulation is always sub-dominant to the observed SB (panel (b) in Figure~\ref{fig:SB_profile}), the simulation is brightest in the South where the observed profiles show a depression in SB. We note that in the jet-aligned sectors, the average SB of the innermost radial bin is $\sim$20\% less than that in the off jet-axis sectors. However, the AGN simulation tends to favor more photons in the jet-aligned sectors, with $\sim$26\% more photons in the North and South compared to the East and West. \textit{Chandra}'s PSF at sub-arcsecond scales is subject to a hook feature which is captured in the \texttt{ChaRT}-\texttt{MARX} simulation. Because the simulation PSF is asymmetric, a simple re-alignement of the simulation is insufficient to eliminate the cavities from the AGN-subtracted SB. 

Our emissivity profiles, temperatures, and metallicities are all measured in the same radial bins/ sectors. Thus, we are able to use the temperatures and metallicities to determine the number density of X-ray emitting electrons $n_e$ from the emissivity profiles. In this way, we obtain profiles of $n_e$ in each sector separately.

\subsection{Contour Binning} \label{sec:cont_bin}

The large signal-to-noise afforded by our deep observations allows us to produce maps tracing the large-scale temperature and metallicity structure in M84. For this task, we use the contour binning method presented in \cite{Sanders2006} and made possible through the \texttt{contbin} software package \citep{Sanders2016}.

Contour binning groups adjacent pixels of similar surface brightness to achieve a requested signal-to-noise ratio. The method groups gas expected to be spectrally similar, allowing us to extract spectra with high signal-to-noise. Thus, contour binning produces accurate temperature maps of spatially-resolved extended sources with non-smooth surface brightness distributions. 

We use a signal-to-noise ratio of 32 and set the smooth signal-to-noise parameter to 20. Because of M84's $\mathcal{H}$-shaped emissivity distribution, contours tend to be elongated along the filaments, connecting regions which are too spatially separated to be causally connected. We thus constrain the shape of the contours using \texttt{contbin}'s \texttt{constrainval} parameter, which we set to 1.2. The spectra extracted from the regions defined by \texttt{contbin} are fit using the spectral model defined in \S\ref{sec:galactic_model}; however, we do not include an XRB/AV/CV component in our fits as this component tends to be negligible on the kiloparsec (kpc) scales relevant for the maps.

\section{Results} \label{sec:results}

In this section, we present profiles of temperature, metallicity, and deprojected density measured in four separate sectors---two aligned with the jet axis and two anti-aligned (Figure~\ref{fig:SB_profile}a). We use these measurements to calculate the accretion rates $\dot{M}_{\rm B}$ and efficiencies $\eta$ for Bondi accretion in each of the sectors, where $L_{\rm Jet} + L_{\rm X} = \eta \dot{M} c^2$ and $L_{\rm X}$ is the X-ray luminosity of the AGN. Our main results are summarized in Table~\ref{table:results}. These measurements allow us to explore the large-scale structure of the accretion flow and compare timescales which dictate the flow's dynamics, namely the cooling time $t_{\rm cool}$ and inflow time $t_{\rm inflow}$. We conclude this section with maps of temperature, metallicity, and pseudo-pressure to connect the small-scale physics of accretion with the kpc-scale structure of the X-ray halo. 

\subsection{Density Profile} \label{sec:density_profile}

The top panel of Figure~\ref{fig:temperature_profile} displays the density profile for each sector. Density increases monotonically toward smaller radii in all sectors from $7''$ to $2''$. The scaling of density with radius provides a direct comparison between our data and the theoretical prediction from the adiabatic Bondi solution. We can model the observed density profile as a simple power law,
\begin{equation}
    n_e (r) = n_{e,0} \left( \frac{r}{R_{\rm B}} \right)^{\alpha},
\end{equation}
where $n_{e,0}$ is the number density at the Bondi radius, $r = R_{\rm B}$. Using a Markov chain Monte Carlo \citep[MCMC;][]{Foreman-Mackey2013} method, we fit this power law model to the density data for each sector. We omit the innermost data points at $1''$ in the North and South sector fits as these points are strongly affected by the presence of Bondi radius-scale cavities. Values of $\alpha$ are shown in Table~\ref{table:results}.

\subsection{Temperature Profiles} \label{sec:results_profiles}

\begin{figure}
\centerline{
\includegraphics[width=0.5\textwidth]{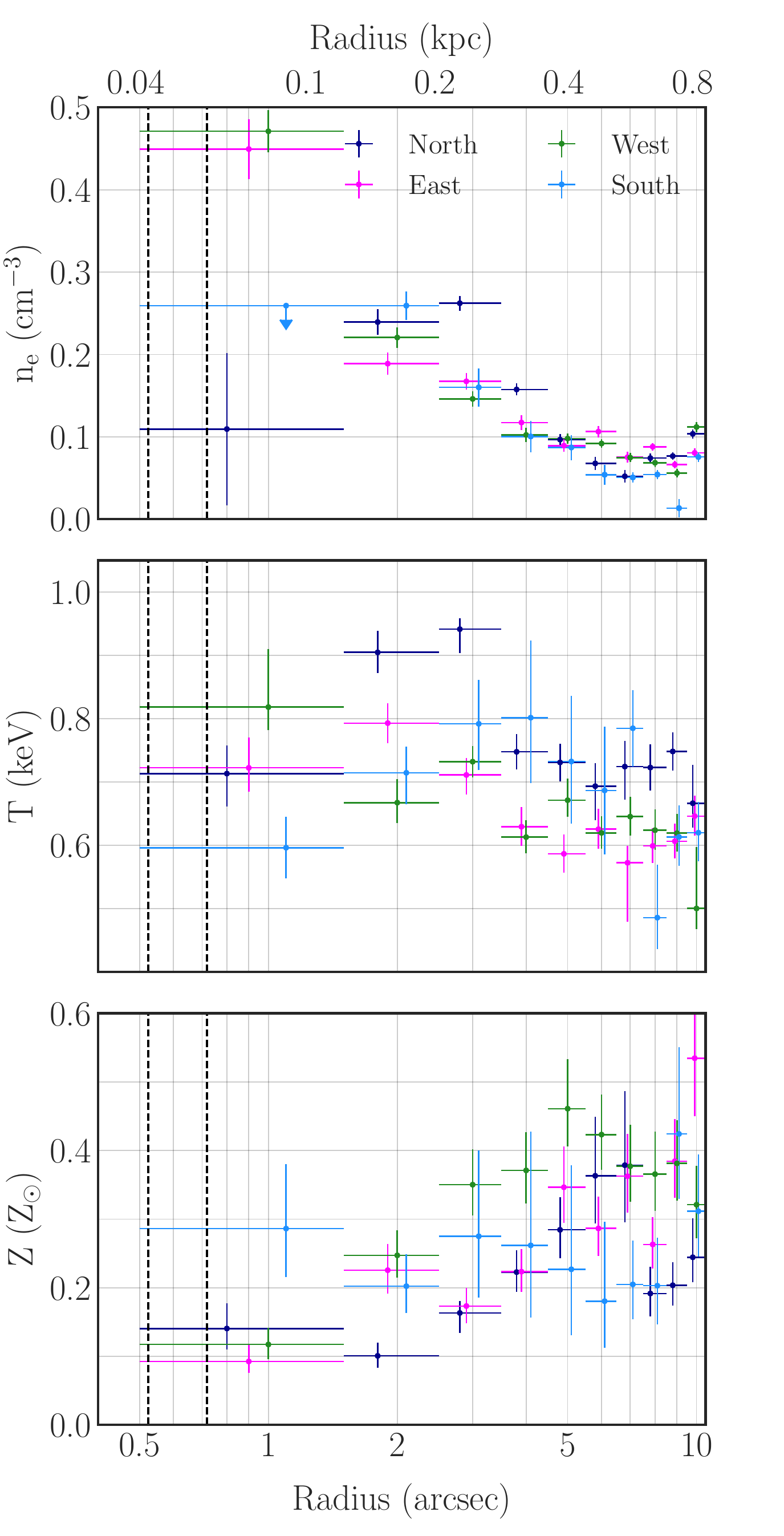}
}
\caption{Deprojected electron number density $n_e$, gas temperature $T$, and iron-peak element metallicity $Z$ as a function of radius for all four sectors. Dashed lines indicate the range of our measured Bondi radii. The presence of cavities in the South sector precludes a measurement of $n_e$ for the innermost point; however, because the emissivity profile drops sharply at this radius, we use the data point at $2''$ as an upper limit.}
\label{fig:temperature_profile}
\end{figure}

We see evidence for shock heated gas in the temperature profile of the North sector (middle panel of Figure~\ref{fig:temperature_profile}). A jump from $T~=~0.75~\pm~0.03$~keV at $4''$ to $T = 0.94^{+0.02}_{-0.04}$ keV at $3''$ in the North sector points to the influence of the radio jet. However, this feature appears to be exceptional rather than commonplace. We find that the temperature profiles are relatively flat with radius, increasing only gradually inward from 0.6 keV at 800 pc to $0.7-0.8$~keV at 100 pc. 

Temperatures at the innermost points, with radial error bars crossing through the Bondi radius, show substantial scatter from $0.6-0.8$~keV. We see that temperature in the South begins to decrease inward at the radii where M84's quasi-spherical central halo begins; gas may be cooling more efficiently in these denser regions. However, we note that the South sector, especially the innermost point, is certainly affected by cavities.

Note that points obtained in the South sector lack the constraining power of the other sectors due to a clear point source throughout much of the region. In all but the South sector, there are at least 700 source counts per radial bin; however, from $3''-6''$ in the South, the number of counts drops below 200. Thus, while the temperature of the innermost point in the South sector may be reliable (although certainly a cavity is present), the decreasing trend in temperature may not be physical. Instead, based on the other sectors, one may reasonably conclude that temperature is relatively constant over the inner kpc, with a gradual increase toward Bondi radius scales. 

\subsection{Bondi Accretion Rate} \label{sec:accretion_rate}

The adiabatic Bondi accretion rate is given by,
\begin{equation} \label{eq:Bondi_accretion_rate}
    \dot{M}_{\rm B} = 0.012 \: \left( \frac{T}{\mathrm{keV}} \right)^{-3/2} \left( \frac{n_e}{\mathrm{cm}^{-3}} \right) \left( \frac{M_{\mathrm{BH}}}{10^9 \: \mathrm{M}_{\odot}}\right)^2 \: \mathrm{M}_{\odot} \mathrm{yr}^{-1},
\end{equation}
where we have measured $n_e$ and $T$ at the Bondi radius \citep{Rafferty2006}, and we use the most recent measurement of the SMBH mass from \cite{Walsh2010}, $M_{\rm BH} = 8.5_{-0.8}^{+0.9} \times 10^8 M_{\odot}$. 

Central values for $\dot{M}_{\rm B}$ are calculated through Equation~\ref{eq:Bondi_accretion_rate} and the data in Table~\ref{table:results}. Our method for computing errors is described in Appendix B. We choose to use a Monte Carlo method, drawing samples from distributions of $n_e$, $T$, and $M_{\rm BH}$ and applying Equation~\ref{eq:Bondi_accretion_rate}. The 1$\sigma$ errors on $\dot{M}_{\rm B}$ are then the 16th and 84th percentiles of the resulting distribution.

Cavities formed by the radio jet and uncertain subtraction of the AGN PSF have a significant impact on measurements of the Bondi accretion rate. Averaging together the East and West sectors to yield $\dot{M}_{\rm B} = (5.94 \pm 0.94) \times 10^{-3} \: M_{\odot} \rm{yr}^{-1}$ perpendicular to the jet, we find that the accretion rate in the North sector parallel to the jet, $\dot{M}_{\rm B} = 1.57^{+1.03}_{-1.01} \times 10^{-3} \: M_{\odot} \rm{yr}^{-1}$, is discrepant at the level of $4.6 \sigma$. Given the dearth of photons in the South sector, the upper limit obtained from this sector is likely a vast over-estimate, with the true inferred $\dot{M}_{\rm B}$ lying even below that in the North sector. These discrepancies point not only to the difficulty of measuring $\dot{M}_{\rm B}$, but also the importance of carefully accounting for cavities, which will systematically suppress the measured $\dot{M}_{\rm B}$. Traditional methods which assume spherical symmetry to compute a deprojected $n_e$ in a full annulus around the AGN rather than in sectors are possibly under-estimating the true Bondi accretion rate.

We close this section by noting that recent measurements by the Event Horizon Telescope \citep[EHT;][]{EHT2019_M87_Shadow} may indicate that the gas dynamical measurement of M84's SMBH mass is an \textit{underestimate} of the true value. The EHT employs an emission modeling technique for assessing SMBH masses which, in the case of Sgr A$^{*}$ \citep{EHT2022_SgrA_Mass} yields a value completely consistent with stellar dynamical measurements \citep{Ghez2008,Gillessen2009}. However, when applied to M87$^{*}$, the EHT measurement \citep{EHT2019_M87_Mass} is discrepant with previous gas dynamical measurements \citep{Walsh2013}. Thus, the \cite{Walsh2010} measurements of M84's SMBH and subsequently our measurements of the Bondi radius and Bondi accretion rate may also be underestimates of the true values. 

\subsection{The Inefficiency of Bondi Accretion} \label{sec:inefficiency}

Using the central values and distributions of $\dot{M}_{\rm B}$, we can compute the efficiency of Bondi accretion. We define this efficiency factor through $L_{\rm Jet} + L_{\rm X} = \eta \dot{M} c^2$, where our combined fit to the AGN gives an X-ray luminosity $L_{\rm X} = 1.6_{-0.14}^{+0.15} \times 10^{39}$ erg/s. The jet power $L_{\rm Jet}$ is obtained by measuring the enthalpy of M84's cavities assuming they are in pressure equilibrium with their surroundings, and dividing by the characteristic timescale of the bubbles, either the sound crossing time or buoyancy timescale. Using this method, \cite{Russell2013} found the jet power to be $L_{\rm Jet} = 1.1_{-0.4}^{+0.9} \times 10^{42}$ erg/s. 

For determining the errors in $\eta$, we assume dimidiated Gaussians for $L_{\rm X}$ and $L_{\rm Jet}$, but instead of assuming distributions for $\dot{M}_{\rm B}$, we use the distributions computed in \S\ref{sec:accretion_rate}. The results are shown in Table~\ref{table:results}. 

While typically $\sim$10\% of the $\dot{M} c^2$ power is released by gravitational infall through an accretion disk, Bondi accretion onto M84's SMBH is far less efficient, with $\eta$~$\sim$~$10^{-6}$ in the East and West sectors unaffected by cavities. These low efficiencies imply that M84 hosts a radiatively inefficient accretion flow (RIAF). A similar conclusion follows from the Eddington ratios for Bondi accretion. Using the definition $L_{\rm B} = \dot{M}_{\rm B} c^2$, and the Eddington luminosity for the SMBH \citep{Russell2013},
\begin{equation}
    L_{\rm Edd} = 1.26 \times 10^{47} \left( \frac{M_{\rm BH}}{10^9 M_{\odot}} \right) \: \rm{erg/s},
\end{equation}
we can determine the Eddington ratio for Bondi accretion, $L_{\rm B}/L_{\rm Edd}$. In all sectors, M84's AGN displays Eddington ratios around a few~$\times 10^{-4}$. When the Eddington ratio is based on the true accretion rate onto the hole $\dot{M}$, the value is much lower, $\eta L_{\rm B}/L_{\rm Edd} \sim 10^{-10}$. Accretion proceeds not through a thin, radiatively efficient disk \citep{Shakura1973}, but rather via a hot RIAF \citep{Yuan2014}.

\begin{figure}
\centerline{
\includegraphics[width=0.5\textwidth]{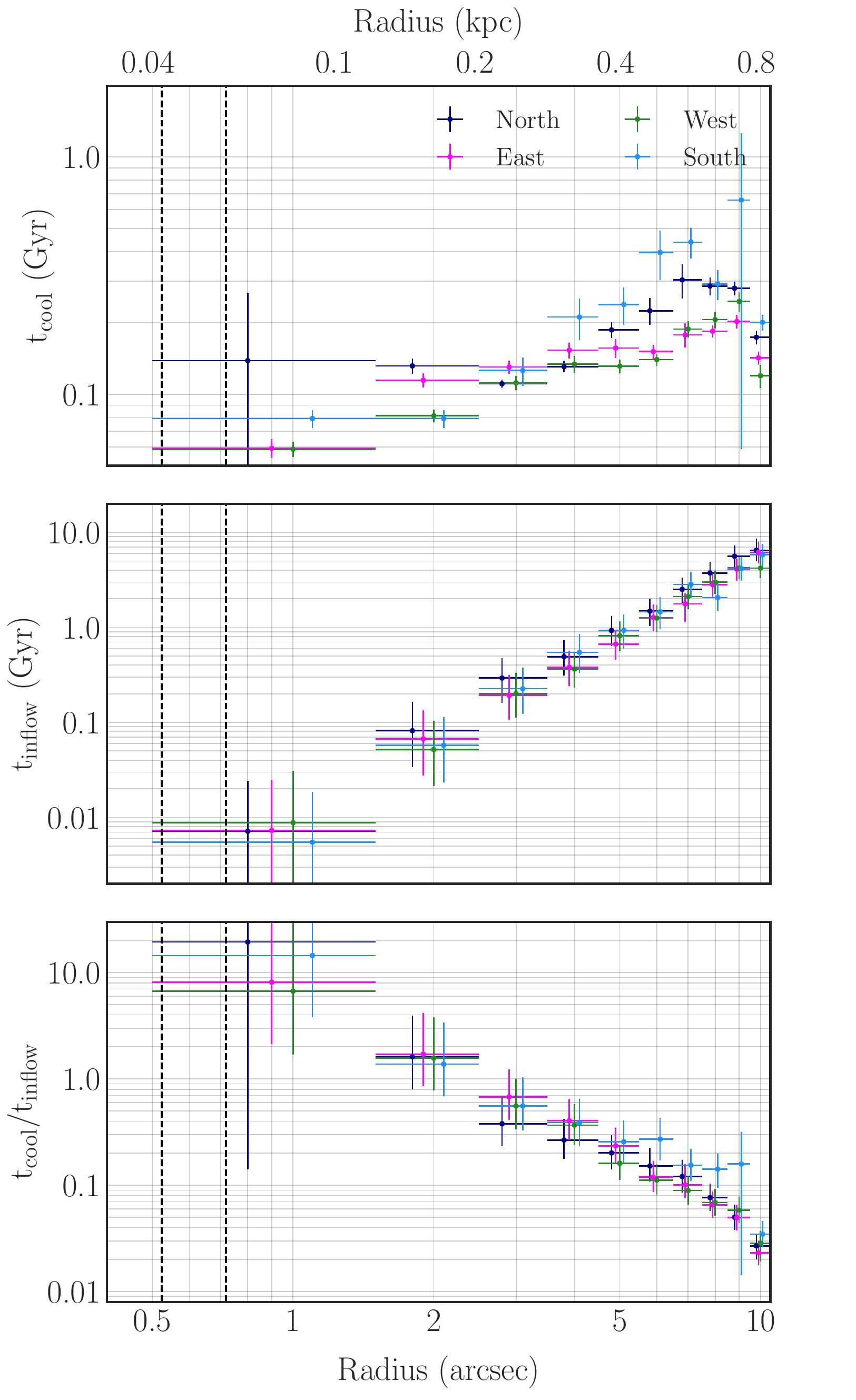}
}
\caption{Profiles of cooling time and inflow time in the inner $10''$ (0.8 kpc). The cooling time $t_{\rm cool}$ gradually decreases inward from 200 Myr at 0.5 kpc to 60 Myr near the Bondi radius (dashed lines). Inflow times show a similar trend; however, the inflow time is $\sim$10 Myr at the Bondi radius in all sectors. The ratio of timescales shows a clear transition above $t_{\rm cool}/t_{\rm inflow} = 1$ at $2''$ as the flow approaches the Bondi radius.}
\label{fig:timescales_profile}
\end{figure}

\subsection{Timescales for Accretion} \label{sec:timescales}

We can better elucidate the structure of the flow by measuring profiles of the relevant timescales for accretion, namely the cooling time $t_{\rm cool}$, free-fall time $t_{\rm ff}$, and Bondi inflow time $t_{\rm inflow}$. The cooling time is the timescale for a gas with thermal energy density $\frac{3}{2} n_e T$ to radiate its energy away, $ t_{\rm cool} = (3/2) n_e T/ n_e^2 \Lambda$. Here, $\Lambda$ represents the cooling function for the X-ray gas and consists of bremsstrahlung continuum as well as significant line cooling in $T \lesssim 1$ keV gas. 

The free-fall time is the dynamical timescale for gas to free fall from a radius $r$ under the gravitational influence of M84's SMBH and dark matter halo. In line with the \cite{Bondi1952} solution, we assume that M84's dark matter distribution is described by a spherically-symmetric \cite{Hernquist1990} profile. We add an additional point mass potential with mass $M_{\rm BH}$ to the dark matter potential, which is used to determine the gravitational acceleration as a function of radius, $g(r)$. The free-fall time can then be computed using the simple expression, $t_{\rm ff} = \sqrt{2r/g}$.

\begin{figure*}
\hbox{
\includegraphics[width=0.5\textwidth]{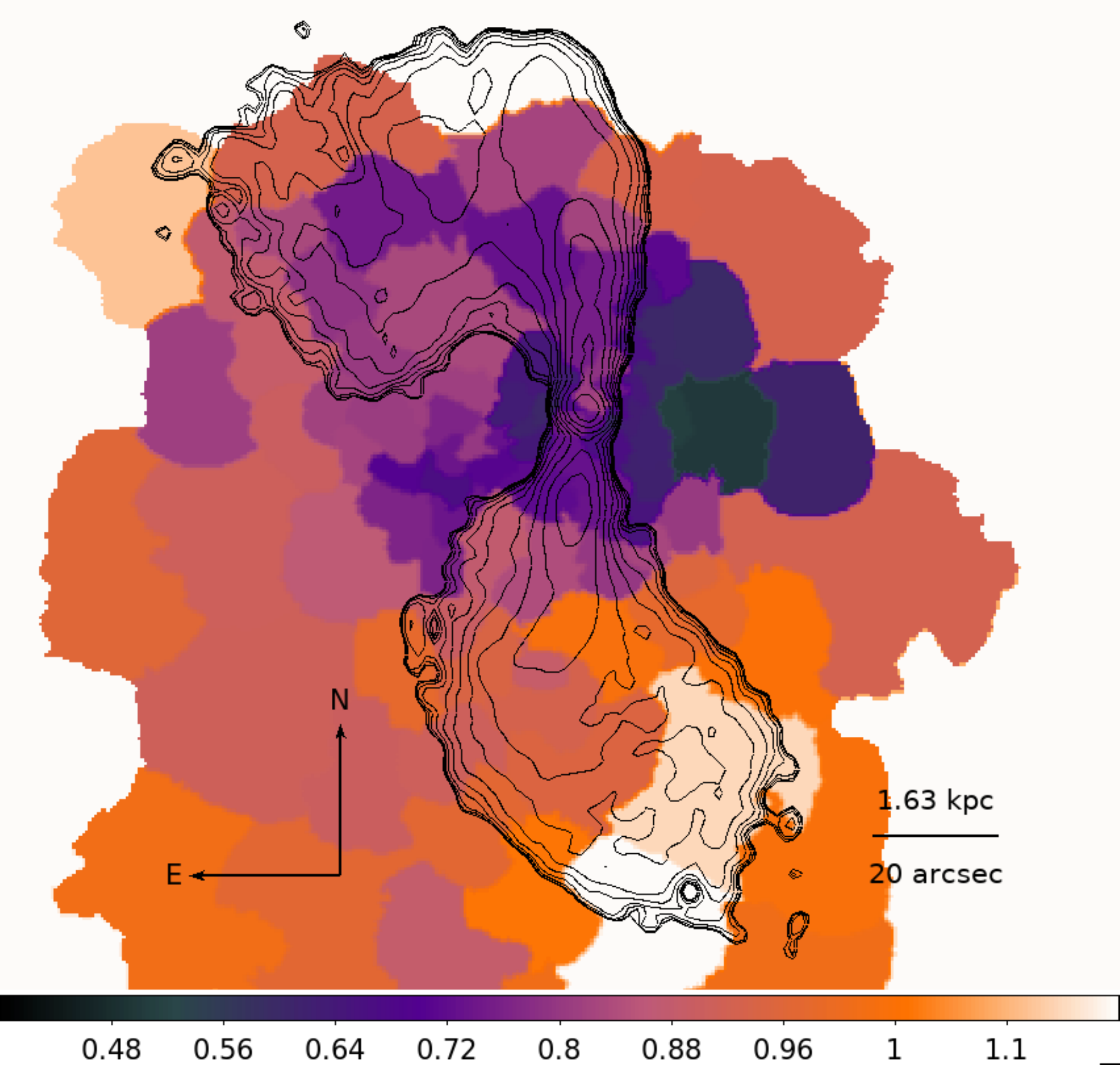}
\includegraphics[width=0.5\textwidth]{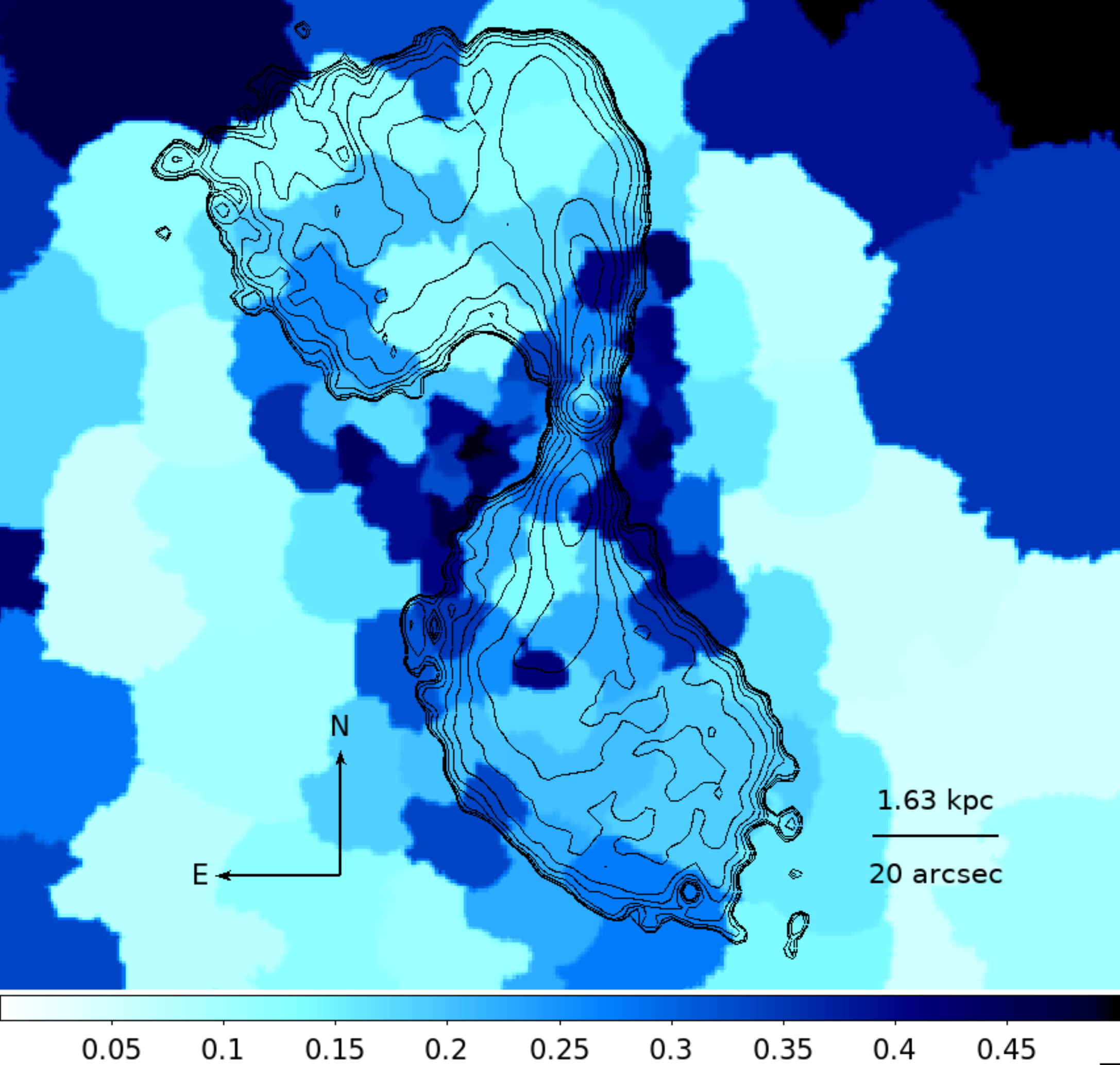}
}
\caption{Left: Temperature map of M84, where brighter colors indicate higher temperature, measured in keV. Right: Metallicity map, where darker colors indicate higher metallicity, measured relative to solar metallicity $Z_{\odot}$. The same $2 \times 10^{-3} - 0.1$ mJy VLA radio contours from Figure~\ref{fig:beautiful_image} are overlaid to emphasize the structure of the $\mathcal{H}$-shaped filaments.}
\label{fig:maps}
\end{figure*}

Traditionally, the ratio $t_{\rm cool}/t_{\rm ff}$ has been used as a probe of thermal instability in galaxy clusters. At the small $\lesssim$ 1 kpc scales where the Bondi flow originates, the free fall time is far too short to be relevant for the structure of the flow (see Discussion). Rather, the dynamical timescale for the accretion flow is the Bondi inflow time, $t_{\rm inflow}$. For steady-state Bondi accreton,
\begin{equation}
    \dot{M}_{\rm B} = 4 \pi r^2 \rho(r) u_r(r) = \rm{constant},  
\end{equation}
where we have introduced the mass density $\rho (r)$ and the radial inflow velocity $u_r(r)$. The mass density is related to our measured number density by assuming quasi-neutrality and introducing the mean-molecular weight $\mu$, which we take to be 0.6 such that, $\rho(r) = 1.15 \: m_p n_e (r)$, and $m_p$ is the proton mass. We can approximate the inflow time as 
\begin{equation}
    t_{\rm inflow} = \frac{r}{u_r(r)} = \frac{4 \pi r^3 \rho (r)}{\dot{M}_{\rm B}}.
\end{equation}
Note that this expression is particularly simple as the number density $n_e$ drops out when substituting in the expression for $\dot{M}_{\rm B}$ (Equation~\ref{eq:Bondi_accretion_rate}),
\begin{equation}
    t_{\rm inflow} \approx 30 \left( \frac{r}{\rm{kpc}} \right)^3 \left( \frac{T}{\rm{keV}} \right)^{3/2} \left( \frac{M_{\rm BH}}{10^9 \: M_{\odot}} \right)^2 \: \rm{Gyr}.
\end{equation}

Figure~\ref{fig:timescales_profile} shows profiles of the timescales. The free-fall timescale is short ($t_{\rm ff}~<~1.2$~Myr) within the inner kpc and does not dictate the gas flow. Rather, cooling is the dominant process from 200$-$800 pc. We emphasize that this flow structure is different from a ``cooling flow'' \citep{Fabian1994} which involves catastrophic levels of star formation and a deluge of cold gas from the cluster ICM onto the galaxy. Our observations probe much smaller scales than would be relevant for cooling flows, which are suppressed by AGN feedback \citep{Fabian2012}. At these small scales, gas slowly condenses and begins to flow inward as the atmosphere cools and loses thermal pressure support.

As material approaches the Bondi radius, the dominant (shorter) timescale becomes the Bondi inflow time $t_{\rm inflow}$. Thermal pressure support loses relevance and the flow begins to experience the influence of the SMBH. In this way, detection of the transition from $t_{\rm cool}/t_{\rm inflow} < 1$ to $t_{\rm cool}/t_{\rm inflow} > 1$ may indicate that we are probing the beginning of the large-scale accretion flow onto the SMBH.

\subsection{Temperature and Metallicity Maps} \label{sec:maps}

The temperature and metallicity profiles presented in Figure~\ref{fig:temperature_profile} provide insight into the structure of the beginnings of the accretion flow feeding the AGN in M84. In this section, we ``zoom out'' from these small, sub-kpc scales to explore the temperature and metallicity structure of the galactic gas. 

Figure~\ref{fig:maps} shows the temperature and metallicity maps produced using the methods described in \S\ref{sec:cont_bin}. The maps are fit using the combined \texttt{VAPEC}+\texttt{APEC} model described in \S\ref{sec:galactic_model} whenever the surface brightness of the region is larger than $10^{-7}$ counts/s/cm$^2$/arcsec$^2$. Otherwise, we use an \texttt{APEC} model since low surface brightness regions are likely dominated by emission from the Virgo ICM rather than from M84's galactic gas. This choice does not have a significant effect on the structure of the maps; the use of an \texttt{APEC}+\texttt{APEC} model for all regions yields similar maps. 

The temperature map shows that the galactic gas is remarkably isothermal, with the vast majority of gas occupying a narrow range of temperatures from 0.5$-$0.8~keV, similar to what is seen in the temperature profiles of the inner kpc around the AGN. There appears to be a large-scale, although weak, temperature gradient, with colder material located West of the radio jet and warmer material sandwiched between the radio lobes. Gas is generally hotter within the radio lobes; however this trend may be due to the fact that the cavities are dominated by dim Virgo ICM emission. In the Northern radio lobe, we see rich temperature structure, with a colder filament bridging through the bubble. This feature may be a projection effect. Instead, a cold filament could be wrapping around in front of or behind the bubble in three dimensional space. 

Colder temperatures trace out filaments; however, the $\mathcal{H}$-shape is far more apparent in the metallicity map in Figure~\ref{fig:maps}. In this map, higher metallicities, with values around 0.3$-$0.5 $Z_{\odot}$ define the $\mathcal{H}$, and metallicity appears relatively symmetric about the radio jet. We see evidence for a drop in metallicity approaching the interior of the galaxy, and this trend is clearly apparent in the metallicity profiles in Figure~\ref{fig:temperature_profile}. Radio cavities appear to clear out the X-ray halo of metal-enriched material; however, rather than pulling high metallicity material into the wake of the bubbles, the jet appears to simply push metal enriched material aside into the $\mathcal{H}$-shaped filaments. 

\begin{figure}
\centerline{
\includegraphics[width=0.5\textwidth]{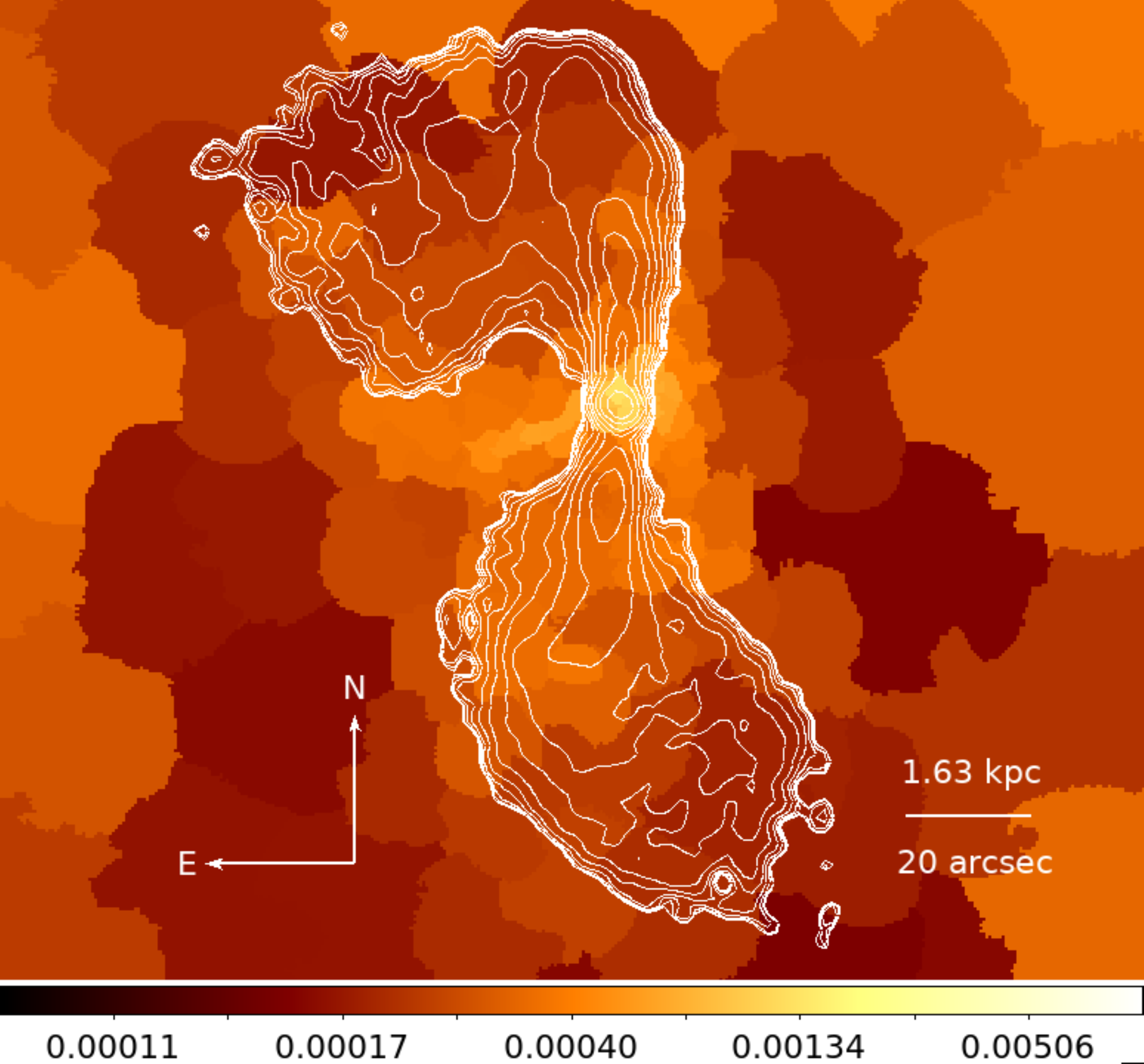}
}
\caption{Pseudo-pressure map, formed by multiplying the temperature map in Figure~\ref{fig:maps} with a map of the square root of area-corrected normalization. The scale shown is logarithmic and spans 2 orders of magnitude. Filaments are not substantially over-pressurized relative to the surrounding cluster.}
\label{fig:pressure_map}
\end{figure}

While jets shape the filaments, feedback is gentle and does not lead to a substantial over-pressurization of the filaments. We can study this process via the pseudo-pressure map (Figure~\ref{fig:pressure_map}). Pseudo-pressure is calculated by multiplying the gas temperature in each region with the square root of the normalization per unit area. For bremsstrahlung emission, which has a weak $T^{1/2}$ temperature dependence, the normalization is proportional to the emission measure, $\int n_e^2 d \ell$, where $d \ell$ is the differential path length through the cluster. 

Pseudo-pressure increases only by a factor 1.5 between the Virgo ICM regions located beyond about $55''$ East of the SMBH and the outer halo of M84. The filaments, which begin $30''$ along the same direction, show another factor of 1.5 increase. Such an increase cannot be explained by the temperature, which instead decreases inward along the same East-pointing ray. The increase is also too steep to be attributable simply to an increase in the path length (and thus emission measure) if we assume M84's galactic gas is distributed quasi-spherically. Instead, gas density is enhanced in the filaments, likely mediated by cooling in the dense, metal-rich gas.
\newpage
The crossbar of the $\mathcal{H}$ can be interpreted as a disk of dense gas around the AGN, viewed in projection. Cooling in the dense disk leads to a collapse into a thin, over-pressurized structure. In this way, rather than being squeezed by the radio lobes, the crossbar may simply be condensing through the cooling and gravitational collapse of a large-scale ($35''$ or $\sim$3 kpc) centrifugally-supported disk. Similarly, features of the pressure map which seem to extend into the radio lobes, such as a ``fish-tail'' like structure visible in the Southeast in Figure~\ref{fig:pressure_map}, can be attributed to filaments wrapping around the radio lobes, viewed in projection. If such filaments are uplifted with the bubbles, cooling may be encouraged, leading to the formation of the dense, metal rich structures clinging to the bubbles. 

\section{Discussion} \label{sec:discussion}

The density, temperature, and metallicity profiles in Figure~\ref{fig:temperature_profile} provide a direct comparison to the spherically-symmetric \cite{Bondi1952} solution. Similarly, our measurements of the profile index $\alpha$ and Bondi accretion rate $\dot{M}_{\rm B}$ (Table~\ref{table:results}) quantify the degree of asymmetry in the flow for sectors aligned with the jet axis and those anti-aligned. 

In this section, we discuss the origins of the observed deviations from the Bondi solution, namely the influence of the jet on the flow. We include a brief discussion of multiphase structure and thermal instability in M84, presenting the entropy profiles based on our temperature and density measurements, and discuss the lack of an observed temperature rise at the Bondi radius. Finally, we close with an analysis of the ``hot blob'' of material noted in \S\ref{sec:results_profiles}.

\subsection{Asymmetry Imposed by the Jet} \label{sec:asymmetry}

Within the 1$\sigma$ error bars, all density profiles are just slightly flatter than $n_e \propto r^{-1}$, which is consistent with findings by \citetalias{Russell2015} in M87. This profile however is completely \textit{in}consistent with the density profile predicted by the \cite{Bondi1952} solution for an adiabatic gas, which instead would predict a density profile of $d \ln{\rho}/ d \ln{r} = -0.373$ at $r = R_{\rm B}$---a 5.5$\sigma$ discrepancy from the ``All'' sectors value in Table~\ref{table:results}. This discrepancy points to the fact that the flow may be strongly influenced by the galactic gravitational potential rather than the SMBH point mass alone \citep{Quataert2000}. 

Both jet-aligned sectors show steeper radial profiles compared with the shallow profiles perpendicular to the jet. Because the points most affected by the presence of cavities were removed when fitting for $\alpha$, the cavities do not account for this steepening in the density profile. Instead, jet-inflated bubbles may entrain dense material from the core of the galaxy in their wakes, buoyantly lifting this gas to larger radii. The result is a dearth of material at the Bondi radius along the jet and density enhancement at larger radii \citep{Churazov2001,Fabian2003}. Alternatively, because the highest densities in the North sector ($n_e = 0.26 \pm 0.009$ $\rm{cm}^{-3}$ at $3''$) are coincident with the ``hot blob'' of shocked gas, the density enhancement may be due to compression in the shocked region itself. 

\subsection{Mechanical Feedback on the Accretion Flow} \label{sec:feedback}

The Bondi accretion rates $\dot{M}_{\rm B}$ aligned and anti-aligned with the jet axis are discrepant at a level $> 4 \sigma$. This discrepancy may be attributed to small-scale cavities formed by radio jets blasting through and clearing out halo gas. However, whether or not this discrepancy indicates that jets modify the true accretion rate of material through the Bondi radius remains an open question. Certainly, cavities complicate measurements of $\dot{M}_{\rm B}$. Spherical symmetry does not apply in the presence of a jet and deprojection is no longer well-posed if surface brightness decreases inward.

Still, the radio jet may have a negligible impact on the accretion flow itself. Relativistic jets with small opening angles can impart substantial energy to the accretion flow via shock heating (see \S\ref{sec:hot_blob}) which impedes gas cooling and introduces further asymmetry to the flow. However, these jets impact a relatively small fraction of the accreting gas. While a large number of simulations have been able to explore self-regulation of AGN in cluster environments \citep{Cattaneo2007,Sijacki2007,Dubois2010,Gaspari2012,Li2014,Prasad2015,Yang2016,Bourne2017}, these simulations lack the dynamic range to study black hole feeding at scales below $R_{\rm B}$. 

Recently, \cite{Ressler2018} and \cite{Ressler2020} demonstrated a calculation of black hole feeding for the RIAF in Sgr A$^{*}$ which evolved the origins of the flow fed by stellar winds down to the black hole horizon. A similar procedure has been undertaken by \cite{Guo2022} for the AGN in M87, with a heating prescription standing in for jetted AGN feedback. In all of these works, angular momentum plays a crucial role. Thermal instability, turbulence, stellar winds, and cloud-cloud or cloud-filament interactions set the angular momentum distribution of accreting gas. High angular momentum gas which is unable to shed angular momentum through collisions of turbulent transport \citep{Narayan2011}, is flung away as it encounters the centrifugal barrier of the SMBH. Yet, low angular momentum gas has the possibility of settling into the observed accretion flow. 

These works predict a suppression of the Bondi accretion rate with the scaling $\dot{M} \sim (r/R_{\rm B})^{1/2} \dot{M}_{\rm B}$. M84's $M_{\rm BH} = 8.5^{+0.9}_{-0.8} \times 10^8 \: M_{\odot}$ black hole has an innermost stable circular orbit (ISCO) with a radius $R_{\rm ISCO} \approx 7.6 \times 10^{14} \: \rm{cm}$ (assuming no black hole spin). For the Bondi radius based on all sectors, $R_{\rm B} (\rm{All}) = 48.1^{+5.3}_{-4.7} \: \rm{pc}$, the predicted accretion rate at the ISCO using the scaling inferred from simulations is $\dot{M} = (R_{\rm ISCO} / R_{\rm B})^{1/2} \dot{M}_{\rm B} \approx 8.5 \times 10^{-6} \: M_{\odot} \rm{yr}^{-1}$. If the flow liberates $\eta \sim 10\%$ of the $\dot{M} c^2$ energy which reaches the hole, the inferred power is $L_{\rm ISCO} \approx 5 \times 10^{40} \: \rm{erg/s}$. This power is well short of the Gyr-averaged jet power $L_{\rm Jet} = 1.1^{+0.9}_{-0.4} \times 10^{42} \: \rm{erg/s}$. Thus, if the $r^{1/2}$ scaling obtains in M84, there must be additional sources of accreting gas beyond the hot phase material inferred from X-ray observations alone.

Understanding the interaction between jets and the accretion flows powering them remains an open problem. Self-consistently evolving the sub-parsec scales responsible for launching jets with the $\sim$50 pc scales of the Bondi radius requires resolving gas thermodynamics, inflows, and outflows over 5 orders of magnitude in scale. We expect that a combination of increased computational power and deep observations of molecular gas, enabled by observatories like the Atacama Large Millimeter Array (\textit{ALMA}), will serve to better elucidate how jets and bubbles affect the distribution of mass and angular momentum in the gas fueling RIAFs in massive elliptical galaxies. 

\subsection{Cold vs. Hot Mode Accretion} \label{sec:multiphase_gas}

Large-scale accretion at scales comparable to the Bondi radius can be broadly divided into two classes, similar to those invoked in the galaxy formation community \citep{Keres2005}: cold mode and hot mode accretion. Bondi accretion of $\sim$keV X-ray gas represents the hot mode of accretion. As we have shown by our measurements of density and temperature at the Bondi radius, Bondi accretion alone is more than sufficient to power the central AGN in M84. However, if multiphase gas, particularly components much colder than what we are studying in the X-rays, is present, the cold mode of accretion may be equally if not more important. 

In cold mode accretion, thermally unstable \citep{Field1965} gas cools, condenses, and precipitates out of the hot medium, forming dense structures such as ``clouds'' (or ``blobs'') and filaments. As long as these cold structures possess a minimal amount of angular momentum, or can shed angular momentum via cloud-cloud, cloud-filament, etc. collisions, they can chaotically ``rain down'' onto the central SMBH, providing a gas supply even in excess of that provided by Bondi accretion alone \citep{Pizzolato2005,Gaspari2012}.

While this picture of accretion is straightforward in principle, in practice, a number of challenges remain. In cluster environments, buoyancy acts to negate thermal instability, at least at the level of linear theory \citep{Defouw1970,Cowie1980,Nulsen1986,Balbus1989}. Idealized nonlinear simulations with heating and cooling globally balanced, as carried out by \cite{McCourt2012} in plane parallel geometry, \cite{Sharma2012} in spherical coordinates, and with jetted feedback as in the simulations by \cite{Gaspari2012}, argue that the existence of multiphase gas depends sensitively on the minimum of the cooling to free-fall time ratio, $t_{\rm cool}/t_{\rm ff}$. Subsequently, simulations by \cite{Li2014} and \cite{Meece2015}, as well as observational efforts by \cite{Voit2015} and \cite{Voit2015_Nature} have further solidified the importance of the ratio of these timescales in the literature.

\begin{figure}
\centerline{
\includegraphics[width=0.5\textwidth]{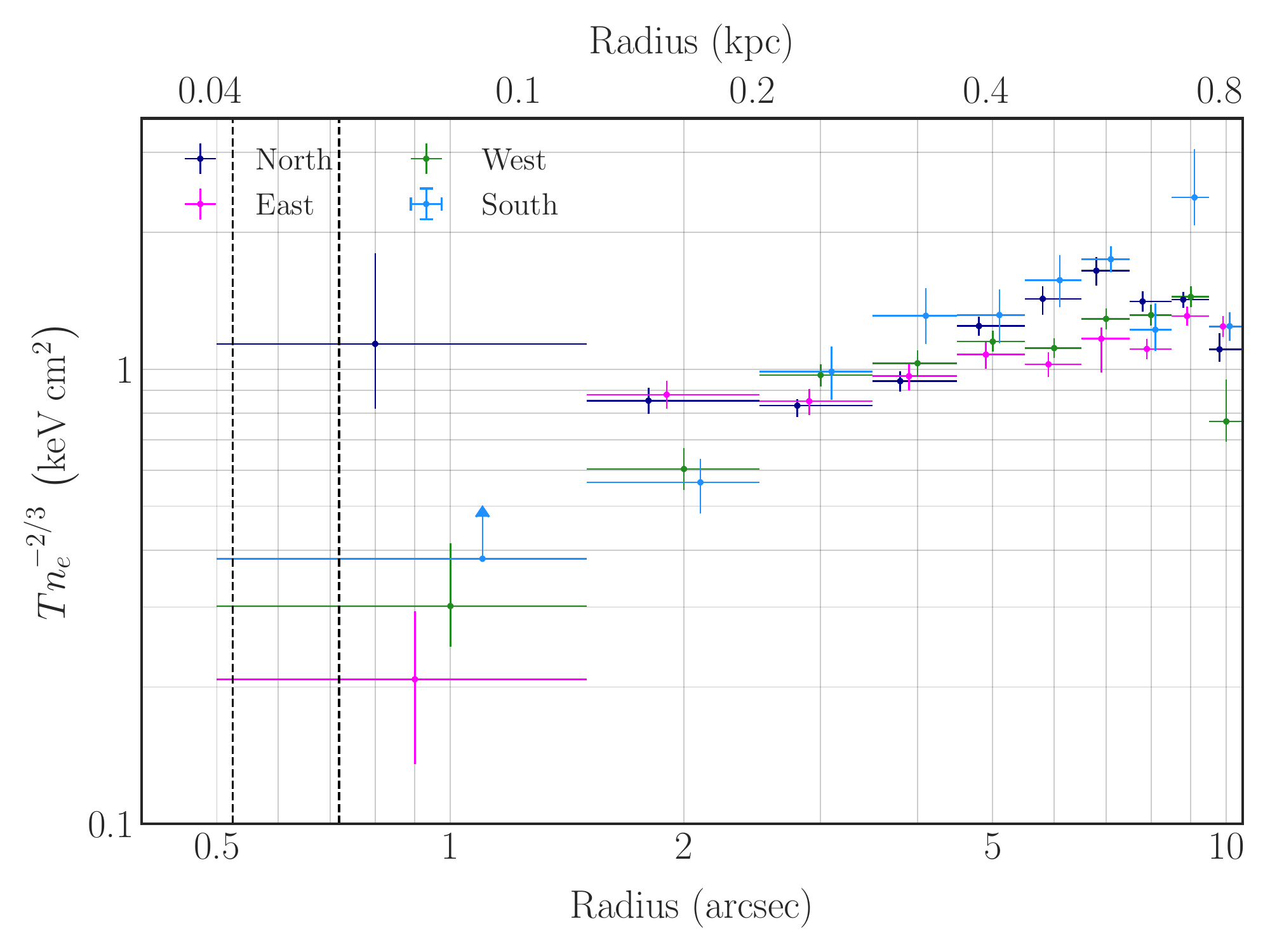}
}
\caption{Entropy profiles of all 4 sectors. There is a clear inward decrease in entropy within the inner kpc of M84, which should indicate a dearth of multiphase gas based on the $t_{\rm cool}/ t_{\rm ff}$ ratio at these small radii. }
\label{fig:entropy_profile}
\end{figure}

\cite{Voit2017} adds motivation for the minimum $t_{\rm cool}/ t_{\rm ff}$ ratio in clusters, arguing that the ratio sets a critical slope for the entropy profiles in clusters. For entropy profiles steeper than this threshold, multiphase gas cannot precipitate from the hot phase since it is subject to buoyant oscillations and thus strong buoyancy damping. Yet, when the slope is flattened by an injection of high entropy material into the center of the cluster via feedback, thermal instability can proceed and cold mode accretion is once again relevant. In this way, the minimum $t_{\rm cool}/ t_{\rm ff}$ ratio alone is not the only crucial parameter in clusters. Rather, this ratio must be compared to the entropy gradient to predict the presence of multiphase gas. 

We note that while there is some observational support for these models, the idea of a critical $t_{\rm cool}/ t_{\rm ff}$ ratio setting the conditions for the formation of multiphase gas is by no means settled physics. Buoyantly rising bubbles may stimulate cooling and multiphase gas formation via adiabatic uplift \citep{McNamara2016}, as may be indicated by our temperature, metallicity, and psuedo-pressure maps (Figures~\ref{fig:maps}-\ref{fig:pressure_map}). In addition, observational biases may over-emphasize the importance of $t_{\rm cool}/ t_{\rm ff}$ \citep{Hogan2017, Pulido2018}. Our work is focused on AGN fueling at the Bondi radius. Thus, rather than wade headlong into rather subtle questions of thermal instability in galaxy clusters, we present a simple test for multiphase gas based on the measured entropy profiles in M84's X-ray halo. 

Figure~\ref{fig:entropy_profile} presents the radial entropy profiles in each of the four sectors. Following \cite{Voit2017}, we define the dimensionless entropy gradient as $\alpha_K \equiv \boldsymbol{r} \cdot \nabla \ln{(K)}$, where $K \equiv T n_e^{-2/3}$ is the gas entropy in units of keV cm$^2$. Equation 22 of \cite{Voit2017} provides a condition on the entropy gradient for nonlinear condensation,
\begin{equation} \label{eq:entropy_gradient}
    \alpha_K < \alpha_{K,\rm{crit}} \equiv \frac{3 (2 - \lambda)^2}{40} \left( \frac{t_{\rm ff}}{t_{\rm cool}} \right)^2,
\end{equation}
where $\lambda \equiv d \ln{\Lambda}/ d \ln{T}$ parameterizes the cooling function $\Lambda$, and for the relevant cooling mechanisms in clusters lies in the range $-1 \lesssim \lambda \lesssim 0.5$. Thus, entropy profiles steeper than the critical value $\alpha_{K,\rm{crit}}$ should result in gradually damped buoyant oscillations of cooling gas, rather than the condensation necessary to fuel cold mode accretion. Because of the short free fall times so close to the Bondi radius ($t_{\rm ff} \sim$ 0.2 Myr) and comparatively long cooling times ($t_{\rm cool} \sim 0.1$ Gyr; Figure~\ref{fig:timescales_profile}) at these small scales, the ratio $t_{\rm ff}/ t_{\rm cool} \sim 2 \times 10^{-3}$ implies that the critical entropy gradient is essentially flat. Figure~\ref{fig:entropy_profile} indicates that in all sectors, the entropy gradient is far too steep to admit condensation and the formation of multiphase gas. If multiphase gas is in fact present, this material must have been sent toward the Bondi radius from much larger scales.

While we searched for multi-temperature gas in the central kpc as an indication of gas cooling out of the ionized phase of the X-ray emitting plasma, we were unable to find evidence of a second temperature component in the X-ray band. Adding in a second \texttt{VAPEC} component provided no constraint on a second temperature. The likely reason is that \textit{Chandra} can only distinguish temperatures separated by $\sim 0.5$ keV in energy space. Because M84's gas is cold ($0.6-0.7$~keV), a colder component would appear at $0.2-0.3$ keV, below the detector's sensitive energy range. A hotter component may be detectable; however, if the AGN was over-subtracted rather than under-subtracted, this potentially weak signal may be lost. Thus, consistent with the conclusion from the entropy profiles, we find no evidence for the presence of multiphase X-ray emitting gas. 

We note that the strict criterion presented in Equation~\ref{eq:entropy_gradient} may not be applicable to Bondi radius scales. The central assumption underpinning the importance of $t_{\rm ff}/t_{\rm cool}$ in cluster environments is that heating and cooling is globally balanced. While such a balance may apply globally within the Virgo Cluster, locally, at the small sub-kpc scales probed in our analysis, heating cannot offset cooling. AGN jet energy is thermalized on 10s of kpc length scales comparable to or larger than the bubbles, via weak shocks and sound waves \citep{Fabian2003, Sanders2007, Graham2008, Fabian2017, Bambic2019}, turbulence driven by g-modes \citep{Churazov2002, Churazov2004, Zhuravleva2014, Zhang2018}, mixing of high entropy bubble material with cluster gas \citep{Hillel2016, Hillel2018}, cosmic rays \citep{Guo2008, Pfrommer2013, Ruszkowski2017, Ehlert2018, Yang2019, Kempski2020}, etc. Near the Bondi radius of M84 where the cooling time is $\sim$0.1 Gyr, these heating mechanisms operate inefficiently. 

For this reason, the traditional comparison of the free-fall and cooling timescales should give way to a comparison of the cooling and Bondi inflow timescales (see \S\ref{sec:timescales}). Figure~\ref{fig:timescales_profile} indicates a transition from a ``cooling-dominated'' flow to an ``inflow-dominated'' flow at scales comparable to $R_{\rm B}$. In this way, we see that the accretion flow around the Bondi radius should not be regarded as a static equilibrium defined by the interplay of AGN heating and radiative cooling, but a dynamic inflow of material under the influence of the SMBH.

\subsection{Metallicity Structure} \label{sec:metallicity}

When the iron-peak element metallicity is free to vary in the fit, we see a clear metallicity gradient. Metallicity decreases inwards in all sectors, with the exception being the cavity-affected point in the South. This gradient (referred to in the literature as a central abundance drop) is common in cluster environments and has been observed in more than 8 objects \citep{Panagoulia2013, Panagoulia2015, Lakhchaura2019, Liu2019_abundance}. Though dust from old stars should be increasing the central metallicity, if this dust is locked into filaments, it can be lifted buoyantly to larger radii by jet-inflated bubbles. Thus, the same processes which shape the density gradients along the jet axis may be responsible for the metallicity gradient.

While buoyancy may explain some of the metallicity gradient along the jet axis, the challenge remains to explain the central abundance drops in sectors perpendicular to the jet axis. Without a modulation of the SNe~Ia rate with radius in the galaxy, this gradient is difficult to account for. Rather than arising from a physical process, the drop in metallicity may point to an unresolved second temperature component and thus, gas cooling en route to a multiphase structure. While we found no evidence for such multi-temperature structure (\S\ref{sec:multiphase_gas}), future, deeper observations free of the constraints on soft energy response which afflict \textit{Chandra} are necessary to tease out the existence of this cooler material. 

We close by noting that absorption, specifically ``intrinsic'' absorption due to interlaced cold and hot phase gas may be obstructing our view of the gas cooling which is responsible for fueling M84's AGN. Such a ``hidden'' cooling flow \citep{Fabian1994, Fabian2022} may be present within M84. Indeed, there is evidence from XMM-\textit{Newton} observations that an intrinsic absorption model may describe M84's galactic gas (Fabian et al. 2022 in prep.). However, high spectral resolution, far beyond what can be achieved by \textit{Chandra}, is required to tease out the parameters of this model. Thus, XMM-\textit{Newton} observations, which probe much larger scales than \textit{Chandra} (comparable with the extent of M84's $\mathcal{H}$-shaped filaments) are unable to constrain an accretion rate for a ``hidden'' cooling flow at Bondi radius scales. 

\subsection{No Temperature Rise at the Bondi Radius} \label{sec:temperature_jump}

We find no evidence for a temperature rise approaching the Bondi radius, a phenomenon that has been proposed as evidence for the transition from the galactic potential to that of the SMBH. This conclusion may be a consequence of the changing metallicity, which decreases by nearly a factor of 4 over the inner few hundred pc in all but the South sector (although the cavity and limited numbers of counts may be playing a role). 

When we fix the metallicity to the radially averaged value ($\sim$0.3~$Z_{\odot}$), we see signs of a temperature jump, with the innermost points in the East and West sectors reaching 1.5 and 1.2~keV respectively. While temperature does increase with fixed iron-peak element metallicity, so also does the reduced C-statistic, indicating that a temperature rise may not truly be present. Instead, there may have been an under-subtraction of the AGN which provides an excess of hard photons, enough to over-estimate the temperature when metallicity is not a free parameter. 

The lack of an observed temperature rise may not be surprising. Observations of the temperature profiles in M87 by \citetalias{Russell2015} find a similar absence. In some respects this is to be expected: the analytical Bondi solution at radii comparable to $R_{\rm B}$ shows a relatively flat temperature profile, with the majority of the adiabatic heating occurring at small scales, well below the Bondi radius. 

However, the notable absence of the temperature rise may be a result of biases inherent to observations of multiphase gas. Indeed, \cite{Guo2022} have recently performed simulations of Bondi accretion from tens of kpc scales down to accretion flow scales well below the Bondi radius which include gas at a wide range of temperatures. They find that their simulations tend to predict a flat emission-weighted temperature profile in the $0.5-7$ keV band, even at scales an order of magnitude below $R_{\rm B}$, where adiabatic heating of hot phase gas becomes significant. Our X-ray observations may be biased by the energy band accessible to \textit{Chandra}, and even future missions which probe scales below $R_{\rm B}$ may similarly never detect a temperature rise.  

\subsection{Shock Heating by the Jet or Nonthermal Emission?} \label{sec:hot_blob}

While there is no evidence for a temperature rise at the Bondi radius, we do see a clear (3.8$\sigma$) temperature increase from $4''$ to $3''$ in the North sector which we refer to as a ``hot blob.'' This temperature increase at $2''-3''$ from the AGN is at the same angular separation from the AGN as a knot of radio emission detected by \textit{VLA} in the 5 and 8.5 GHz bands, and \textit{ALMA} in the 97 and 236 GHz bands (see Knot B in Figures 1-3 of \cite{Meyer2018}). This ``blob'' or ``knot'' of X-ray emission, first detected by \cite{Harris2002}, stands out clearly in X-rays, even with the limited exposure time ($\approx$29 ks) of \textit{Chandra}'s first observation of M84.

The $0.5-7$ keV X-ray spectrum of the region containing the ``hot blob'' is well described (reduced C-statistic of 1.08 and 1.06 for the North Sector at 2$''$ and 3$''$ respectively) by a \texttt{VAPEC} model in our analysis, indicating thermal X-rays. However, the radio emission is far more complicated. Early results favored a synchrotron origin for the emission \citep{Harris2002}; however, \cite{Meyer2018} argue that the radio and X-ray spectra of the knots cannot both be explained by standard models for jet emission. In their analysis, the X-ray emission is modeled as both a power law representing the jet and an \texttt{APEC} component representing the thermal gas. 

Motivated by these works, we re-fit the spectra from the North sector at 2$''$ and 3$''$ using our M84 model (\S\ref{sec:AGN_model}), with the addition of a red-shifted power law (\texttt{zpowerlw}) component meant to represent the X-ray jet detected by \cite{Meyer2018}. When fitting the 2$''$ and 3$''$ North spectra with the extra \texttt{zpowerlw} ``jet'' component, the resulting temperature and metallicity of the \texttt{VAPEC} component are unconstrained (in the case of the 3$''$ point, temperature is constrained but metallicity is not). We thus proceeded to leave the temperature and \texttt{zpowerlw} normalization free in the fit, but fix the \texttt{VAPEC} metallicity to three different values of metallicity: 0.1 (consistent with our measurements), 0.2, and 0.3 solar (consistent with that used by \cite{Meyer2018}). For each metallicity, we scan \texttt{zpowerlw} photon indices from $\Gamma = 1-3$, which we fix in the fit.

This procedure yields improved C-statistics over the 1.08 and 1.06 found initially, in some cases comparable to or better than the fit to the full annulus spectrum including all sectors at 2$''$ and 3$''$. Reasonable photon indices near $\Gamma \approx 2$ yield good fits. As expected, the corresponding \texttt{VAPEC} temperature is lower when the jet component is included; however, rather intriguingly, the fit temperature is lower than all other sectors save that in the West at the same radii. The temperature never exceeds 0.7 keV for all photon indices and metallicities tested. For the metallicity of 0.1 solar consistent with what was found in our profiles (Figure~\ref{fig:temperature_profile}), the fit finds \texttt{VAPEC} temperatures below 0.6 keV for the point at 2$''$. When using the metallicitity of 0.3 solar assumed in \cite{Meyer2018}, the fits settle around $T = 0.65$ keV, which is just below the temperature at a radius of 2$''$ in the East Sector ($T = 0.67^{+0.04}_{-0.03}$ keV). Note that all of these temperatures are well below the $3$ keV thermal model for the X-ray emission proposed as an alternative to synchrotron emission in \cite{Harris2002}.

The takeaway message from this analysis is clear: a model with nonthermal emission from an X-ray jet and colder ($T \approx 0.65$ keV) galactic gas describes the ``hot blob'' as well as a purely thermal emission model of $T \gtrsim$ 0.9 keV gas shock heated by the jet. Because of \textit{Chandra}'s limited spectral resolution, we cannot discriminate between these models. However, given the complexity of M84's jet emission in the radio band and the co-spatial temperature increase observed in our X-ray observations, a deeper study of M84's jet which can harness the full power of our 840 ks data set is merited.  

\subsection{Comparison to Other Measurements} \label{sec:comparison}

Bondi accretion is incredibly inefficient. The Bondi accretion rate $\dot{M}_{\rm B}$ is measured to be a few $\times \: 10^{-3} M_{\odot} \rm{yr}^{-1}$ in each of the sectors in this work. Thus, the accretion flow in M84 need only liberate  $\eta \sim 10^{-6}$ of the $\dot{M}_{\rm B} c^2$ fuel provided by Bondi accretion to power the galaxy's relativistic jets and X-ray AGN. 

We are not the first group to arrive at this conclusion in M84. The first measurement of the Bondi accretion rate can be attributed to \cite{Allen2006}, who found an accretion rate of $\dot{M}_{\rm B} = 8.5^{+8.4}_{-4.1} \times 10^{-3} M_{\odot} \rm{yr}^{-1}$ by measuring the temperature and density of the full annulus around the AGN, i.e. including all sectors at the innermost ``Bondi radius'' point. While this work was in preparation, another measurement of $\dot{M}_{\rm B}$ was performed by \cite{Plsek2022} which leveraged the data from our new campaign, presented here and publicly available. They found $\dot{M}_{\rm B} = 2.4^{+1.9}_{-1.5} \times 10^{-3} M_{\odot} \rm{yr}^{-1}$, again using all sectors. If we compare these values with our ``All'' sector measurement of $\dot{M}_{\rm B} = 3.74^{+1.05}_{-0.89} \times 10^{-3} M_{\odot} \rm{yr}^{-1}$, then there is strong agreement among all published values of the Bondi accretion rate in M84, with notably tighter error bars in the more recent values enabled by an extra $\gtrsim 750$ ks provided by the new campaign. 

\section{Conclusion} \label{sec:conclusion}

We have presented the deepest \textit{Chandra} X-ray observations to date of M84, a jetted elliptical galaxy in the Virgo Cluster. These observations, which comprise over 840 ks of \textit{Chandra} data, have enabled a detailed study of the temperature, density, and metallicity structure of the galaxy, from kiloparsec scales to $\approx 50$ pc scales just inside the Bondi radius of the galaxy's SMBH. New images of M84 have been presented, emphasizing the intricate structure of the soft X-ray filaments, formed into an $\mathcal{H}$ morphology by the action of powerful \citep[$L_{\rm Jet} = 1.1^{+0.9}_{-0.4} \times 10^{42}$ erg/s;][]{Russell2013} radio jets.

Density and temperature measurements obtained through spectra extracted from the innermost $0.5''-1.5''$ bin allowed us to compute Bondi accretion rates for each of 4 sectors around the central AGN. ``All'' sectors are fit together to allow comparison to previous works. The main conclusions of our analysis are as follows:

\begin{enumerate} 

\item Radial profiles of deprojected electron number density $n_e$ are consistent with $n_e \propto r^{-1}$, but slightly flatter (Figure~\ref{fig:temperature_profile} and Table~\ref{table:results}). This profile is in tension at the level of $5.5 \sigma$ with the expectation of Bondi accretion, which predicts $d \ln{\rho} / d \ln{r} = -0.373$ at $r = R_B$ and an $r^{-3/2}$ scaling at $r \ll R_B$.

\item The radial profile indices $\alpha$ are statistically consistent; however, we see that the profiles are steeper along the jet axis than perpendicular to the jet (Table~\ref{table:results}). This violation of spherical symmetry is counter to the assumptions of the Bondi solution.

\item There is a discrepancy in the Bondi accretion rate depending upon which sector is used to measure $\dot{M}_{B}$ (Table~\ref{table:results}). This discrepancy between jet-aligned and mis-aligned sectors is at the level of $4.6 \sigma$, which is statistically significant. While the discrepancy may point to the influence of the jet on the large-scale accretion flow, the disparity likely arises due to the presence of cavities (Figures~\ref{fig:SB_profile} and \ref{fig:temperature_profile}) or uncertainties in modeling the AGN emission at Bondi radius scales (see Appendix A).

\item Temperatures do not vary widely throughout the galaxy (Figure~\ref{fig:maps}) and only increase gradually from 0.6 keV to 0.7 keV over the inner kpc approaching the Bondi radius. The exception is that in the North sector, we see evidence for a temperature increase at points $2''-3''$ from the AGN (Figure~\ref{fig:temperature_profile}). We refer to this feature as a ``hot blob'' of gas. The physical origin of this ``hot blob'' remains an open question. Shock heating by the radio jet or nonthermal emission from knots in an unresolved X-ray jet \citep{Meyer2018} are both plausible explanations; however, \textit{Chandra} lacks the spectral resolution to rule out either of these two models. 

\item We detect no temperature rise at the Bondi radius, consistent with findings by \citetalias{Russell2015} in M87.

\item By comparing the Bondi inflow time $t_{\rm inflow}$ to the cooling time as a function of radius, we observe evidence for a transition from a ``cooling-dominated'' flow to an ``inflow-dominated'' flow at scales of $1''-2''$,  providing support to the conclusion that we have resolved M84's Bondi radius. 

\end{enumerate}

\section*{Acknowledgements}

The data presented in this work was obtained through \textit{Chandra} Proposal \#19800344: ``Fueling and self-regulation of AGN feedback at the Bondi radius of M84'' and is publically available on the \textit{Chandra} archive. We are grateful to the \textit{Chandra} X-ray Center for support, not only with the initial observations and pointing of the telescope, but also with use of the \texttt{ChaRT} tool. This manuscript was improved thanks to the careful reading and suggestions of an anonymous referee. CJB thanks Andy Goulding and Jeremy Sanders for technical support throughout this project. This work benefited from stimulating discussions with Eliot Quataert and Minghao Guo at Princeton and Eileen Meyer at the Texas Symposium on Relativistic Astrophysics. CJB thanks the graduate students of the Institute of Astronomy, Cambridge and Princeton University for their many insights as well. This work is possible through the financial support of the Churchill Foundation of the United States and CJB continues to be supported by a National Science Foundation (NSF) Graduate Research Fellowship.  Early stages of this work were performed at the Multiscale Phenomena in Plasma Astrophysics program at KITP in Santa Barbara, CA, research supported in part by the NSF under Grant No. NSF PHY-1748958. HRR acknowledges support from an STFC Ernest Rutherford Fellowship and an Anne McLaren Fellowship provided by the University of Nottingham. CSR thanks the STFC for support under the Consolidated Grant ST/S000623/1, as well as the European Research Council (ERC) for support under the European Union’s Horizon 2020 research and innovation programme (grant 834203).

\section*{Data Availability}

The \textit{Chandra} data described in this work are available in the Chandra data archive (https://cxc.harvard.edu/cda/). Processed data products detailed in this paper will be made available on reasonable request to the author.

\bibliographystyle{mnras_mwilliams} 
\bibliography{M84.bib}

\section*{Appendix A: Modeling AGN Contamination}

\texttt{ChaRT} and \texttt{MARX} simulations of the AGN require an input energy spectrum, which was obtained by fitting the spectrum extracted from a $1''$ circle about the AGN. While this spectrum is dominated by the AGN ``source,'' it includes ``background'' contributions from M84's galactic gas, the Virgo screen, and unresolved XRB/AB/CV stars. The Virgo and unresolved point source backgrounds were determined in \S\ref{sec:virgo_model} and \S\ref{sec:unresolved_points} respectively; however, the treatment of the underlying galactic gas emission required more care. 

\begin{figure}
\centerline{
\includegraphics[width=0.5\textwidth]{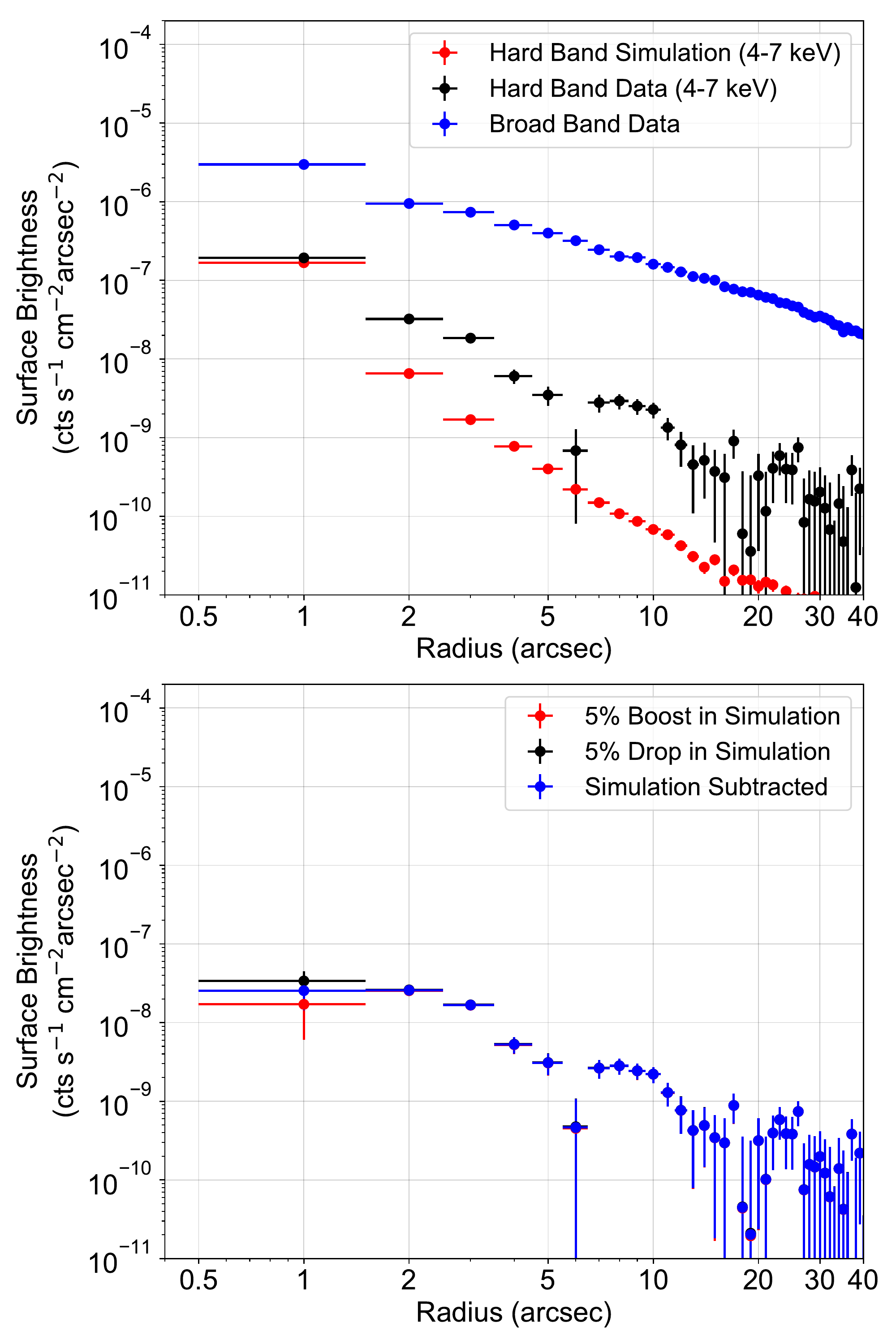}
}
\caption{Top: Hard band ($4-7$ keV) surface brightness profiles extracted from merged event file and \texttt{ChaRT}+\texttt{MARX} AGN simulation, with 7\% boost applied to simulated profiles. Bottom: Subtraction of hard band simulation profile from hard band data profile. The hard band, where only AGN emission and background from the Virgo screen and unresolved point sources remains, is more than an order of magnitude subdominant to the total SB. By applying a 7\% boost to the simulation, we achieve a flattening of the AGN-subtracted profile within the inner $2''$, indicating that the AGN has been properly subtracted from the data. All that remains at PSF scales are negligible contributions from the spatially uniform Virgo screen and unresolved point sources. The point source emission appears to be subdominant given the smoothness of the AGN-subtracted hard band SB from $1''-3''$.}
\label{fig:hard_band}
\end{figure}

To set the background galactic gas component at $1''$, we fit the spectrum of an annulus from $2''-4''$ circumscribing the AGN where AGN emission is negligible. This procedure provided a normalization for the \texttt{VAPEC} background component at 3$''$ from the AGN. Then, we fit the surface brightness (SB) distribution with the assumption of spherical symmetry using a simple power law in radius, assuming that the power law extrapolation from $3''$ to $1''$ accurately described the SB at $1''$ from the central AGN. The ratio of the SB determined at $1''$ and $3''$ from this power law was taken as a ``boost'' factor multiplied on to the previously fit-for \texttt{VAPEC} normalization. In the case of M84, we found this boost factor to be 3.98.

Then, the parameters of the input energy spectrum passed to \texttt{ChaRT} were determined by fitting the full spectrum (source + background) extracted from the $1''$ circle with a \textit{fixed} background \texttt{VAPEC} component and only the \texttt{zpowerlw} parameters (meant to model the AGN source) left free. The normalization of this fixed \texttt{VAPEC} component is simply the normalization of the $2''-4''$ annulus, multiplied by the boost factor (3.98) and corrected by the ratio of the annulus to $1''$ circle areas. Once the parameters for the \texttt{zpowerlw} source component were determined, a clean spectrum including only \texttt{phabs(zphabs(zpowerlw))} parameters was produced and passed to the \texttt{ChaRT} tool as the input spectrum. This source spectrum is representative of the AGN without contributions from the background.

\newpage

Because the normalization of the input spectrum was determined by assuming a model for the galactic gas spectrum, the flux of the simulation may not be an accurate representation of the true AGN. To test the accuracy of the AGN modeling, we choose an energy band where the AGN completely dominates and which is free of the small-scale variations (on scales comparable to the PSF) imposed by bright, lumpy, soft emission from galactic gas. In this case, following \citetalias{Russell2015}, we choose the hard $4-7$ keV band. At these energies, the only contributions to the hard band SB should be from the AGN source, Virgo ICM background (which should be spatially uniform), and unresolved point source background. If point source emission is substantial, the hard band SB profile should display discontinuities and rapid spatial variations on scales of the PSF. 

\newpage

The top panel of Figure~\ref{fig:hard_band} shows a comparison of the hard band profiles for the data (black) and simulation (red). By forward modeling the AGN, we are working to subtract the AGN contribution at Bondi radius scales from the spectra extracted in each sector. The blue points in the lower panel of Figure~\ref{fig:hard_band} show a subtraction of the simulation's hard band SB profile from that of the data. If we were to overestimate the AGN flux and thus over-subtract the AGN from the data, we should expect a drop in hard band SB at $1''$, i.e. the scales of the PSF. Alternatively, if we under-subtract the AGN, we should expect a hard band excess in the bottom panel of Figure~\ref{fig:hard_band}.

We find that our AGN simulation based on modeling the galactic gas background leaves a 7\% excess in the hard ($4-7$ keV) band. Without compensating for this excess, we would under-subtract the AGN and possibly bias our temperature measurements with excess hard AGN photons. Thus, we boost the overall AGN simulation normalization by 7\%, which means that the difference profile in the bottom panel of Figure~\ref{fig:hard_band} flattens at the scales of the PSF. When computing errors on temperature, we do not simply report the statistical uncertainties determined by \texttt{XSPEC}. Rather, we boost the AGN normalization by 5\% (on top of the 7\% compensation) and decrease the normalization by 5\% (shown as the red and black points in the bottom panel of Figure~\ref{fig:hard_band} respectively), to marginalize over uncertainties in the AGN modeling. As a result, errors in temperature and metallicity are larger at the Bondi radius. Finally, because the hard band SB is relatively continuous at PSF scales, we conclude that unresolved point sources are properly accounted for. 

\section*{Appendix B: Ascertaining Errors on $\dot{M}_{\rm B}$ and $\eta$}

For computing errors on $\dot{M}_{\rm B}$, we use a Monte Carlo method, drawing $10^7$ samples from distributions of $n_e$, $T$, and $M_{\rm BH}$ and applying Equation~\ref{eq:Bondi_accretion_rate}. Because of asymmetric error bars in $T$ and $M_{\rm BH}$, we model the distributions of these variables as ``dimidiated Gaussians'' \citep{Barlow2003}, two Gaussians centered on the same mean with different standard deviations above and below the mean based on the 1$\sigma$ upper and lower error bars. For equal positive and negative error bars, the dimidiated Gaussian is equivalent to a normal distribution. We model the underlying distribution of $n_e$ as a log-normal with mean and standard deviation based on the central value and $1\sigma$ error bar respectively. This choice ensures strict positivity of $n_e$ but only has significance for the point in the North---a Gaussian yields a similar error bar for all other points. We take the 1$\sigma$ errors on $\dot{M}_{\rm B}$ to be the 16th and 84th percentile of the resulting distribution.

\clearpage

\end{document}